\documentclass[lettersize,journal]{IEEEtran}
\usepackage{graphicx}
\usepackage{caption}
\usepackage{subcaption}
\usepackage{amsmath,amsfonts}
\usepackage{algorithmic}
\usepackage{array}
\usepackage{textcomp}
\usepackage{stfloats}
\usepackage{url}
\usepackage{verbatim}
\usepackage{graphicx}
\usepackage{booktabs}
\usepackage{makecell}
\usepackage{multirow}
\usepackage{stfloats}
\usepackage{enumitem}
\usepackage{caption}
\usepackage{calc} 
\usepackage{caption}
\usepackage{graphicx}
\usepackage{float} 
\usepackage{tabularx} 
\usepackage{siunitx}  
\usepackage{footnote}
\usepackage{hyperref}
\usepackage{orcidlink} 
\usepackage[backend=biber, style=ieee]{biblatex} %
\def\BibTeX{{\rm B\kern-.05em{\sc i\kern-.025em b}\kern-.08em
    T\kern-.1667em\lower.7ex\hbox{E}\kern-.125emX}}
\usepackage{balance}
\begin{document}
\title{IDU-Detector: A Synergistic Framework for Robust Masquerader Attack Detection}
\author{Zilin Huang\textsuperscript{*}\orcidlink{0009-0001-6940-8085}, Xiulai Li\textsuperscript{*}\textsuperscript{\dag}\orcidlink{0000-0001-7867-2032} \IEEEmembership{Member, IEEE}, Xinyi Cao\orcidlink{0009-0001-0358-8570}, Ke Chen\orcidlink{0009-0004-0850-311X}, Longjuan Wang\orcidlink{0000-0002-1135-6251}, Logan Bo-Yee Liu\orcidlink{0000-0001-7163-5482} \IEEEmembership{Member, IEEE}
\thanks{This research was supported by the Hainan Provincial Natural Science Foundation of China (Grant No. 723QN238  and Grant No. 722RC678) awarded to Xiulai Li.}
\thanks{Z. Huang and K. Chen are with the School of Cyberspace Security (School of Cryptology), Hainan University, Haikou, 570228, China(e-mails: zilin\_huang2024@163.com, chenk@hainanu.edu.cn).}  
\thanks{X. Li is the corresponding author, is with Hainan University, Haikou 570228, China, and Hainan Hairui Zhong Chuang Technol Co Ltd, Haikou 570228, China(Corresponding author e-mail: lixiulai01@hainanu.edu.cn).}
\thanks{X. Cao is with the School of Computer Science and Technology, Hainan University, Haikou 570228, China(e-mail: xinyi\_cao2003@163.com).}
\thanks{L. Wang is with the School of Cyberspace Security(School of Cryptology), Hainan University, Haikou, 570228,China(e-mail:wanglongjuan@hainanu.edu.cn).}
\thanks{Logan B. Liu is with the Hainan Shuyi Technol Co Ltd., Sanya 572000, China(e-mail: by.liu@ieee.org).}
\thanks{* Zilin Huang and  Xiulai Li contributed equally to this work.}
\thanks{$ \dag $ Corresponding author.}
}
\maketitle

 
 

\begin{abstract}
In the current digital age, users store their personal information in corporate databases to access services, making data security and sensitive information protection central to enterprise security management. Given the extensive attack surface, system assets continuously face cyber security challenges such as weak authentication, exploitation of system vulnerabilities, and malicious software. Through specific vulnerabilities, attackers may gain unauthorized system access, masquerading as legitimate users, and remaining hidden. Successful attacks can lead to the leakage of user privacy, disruption of business operations, significant financial losses, and damage to corporate reputation. The increasing complexity of attack vectors is blurring the boundaries between insider and external threats. To address this issue, this paper introduces the IDU-Detector, an innovative threat detection framework that strategically integrates Intrusion Detection Systems (IDS) with User and Entity Behavior Analytics (UEBA). This integration aims to monitor unauthorized access and malicious attacks within systems, bridging functional gaps between existing systems, ensuring continuous monitoring and real-time response of the network environment, and enhancing their collective effectiveness in identifying security threats. Additionally, the existing insider threat datasets exhibit significant deficiencies in both depth and comprehensiveness, lacking sufficient coverage of diverse attack vectors. This limitation hinders the ability of insider threat detection technologies to effectively address the growing complexity and expanding scope of sophisticated attack surfaces. To address these gaps, we propose new, more enriched and diverse datasets that includes a wider range of attack scenarios, thereby enhancing the adaptability and effectiveness of detection technologies in complex threat environments. We tested our framework on different datasets, the IDU-Detector achieved average accuracy rates of 98.96\% and 99.12\%. These results demonstrate the method's effectiveness in detecting masquerader attacks and other malicious activities, significantly improving security protection and incident response speed, and providing a higher level of security assurance for asset safety.
\end{abstract}

\begin{IEEEkeywords}
masquerader attacks, deep learning, intrusion detection, insider detection.
\end{IEEEkeywords}

\section{Introduction}
\IEEEPARstart{I}{n} the modern era, characterized by rapid technological advancement and accelerated digital transformation, enterprises and organizations increasingly depend on complex information systems to manage their critical business operations and sensitive data. Within this context, cybersecurity emerges as a crucial challenge for protecting corporate assets and ensuring the continuity of business operations \cite{r1_1rawal2022identifying,r1_2jo2019mauth}. This challenge extends to various domains, including emerging technologies like federated learning systems and cloud robotics, where data privacy and security are paramount \cite{liu2019federated,liu2019lifelong, liu2021peer}

Internet-borne security threats, prevalent in today's connected world, can be categorized into external and insider threats. Notably, insider threats pose the most harmful risk to corporate and organizational assets. These threats originate from malicious users who possess legitimate access rights. Typically, such individuals are well-acquainted with the organizational information system architecture and sensitive data access \cite{r1_3alotibi2019feasibility, zhang2022authros}, enabling them to circumvent traditional security measures relatively easily. They exploit the trust and permissions granted to legitimate users to perform operations that could cause significant, often irreversible, damage to the organization's infrastructure and data integrity. As a result, insider threat represent a primary challenge to asset security. Conversely, external threats are usually initiated by unauthorized attackers who attempt to gain legitimate access to systems, a process known as intrusion. The successful execution of such attacks often allows attackers to acquire access rights, subsequently elevating their status to that of insiders. This malicious privilege escalation is particularly perilous because once external attackers gain access, they can operate within the network with minimal suspicion, accessing sensitive information and critical systems undetected.

\begin{table*}[bp]
  \centering
  \caption{Cyber Attack Vectors and Their Implications on System Security}
  \fontsize{10}{12}\selectfont
  \label{tab:cyber-attack-vectors}%
  \begin{tabular}{p{1cm}p{4.5cm}p{4.5cm}p{4cm}p{2cm}}
    \toprule[0.8pt]
     \parbox{1cm}{\centering \textbf{Attack Vector}} &
    \parbox{4.5cm}{\centering \textbf{Definition}} &
    \parbox{4.5cm}{\centering \textbf{Impact/Purpose}} &
    \parbox{4cm}{\centering \textbf{Pathway to Obtain Legitimate Access}} &
    \parbox{2cm}{\centering \textbf{Classification}} \\
    \midrule[0.8pt]
    Probe & Attackers use scanning or probing techniques to gather information about the target system or network, such as vulnerabilities, configurations, open ports, and available services. & Preparation for subsequent attacks (e.g., privilege escalation, data theft, DoS attacks). Probing itself does not necessarily compromise the system, but it serves as the initial step for more severe attacks, posing a potential threat to system security. & Probing does not directly obtain access rights but provides pathways for subsequent attacks. &  \makecell[lt]{Potential\\ Intruder} \\
    \midrule[0.8pt]
    U2R & Attackers exploit system vulnerabilities or privilege escalation techniques to elevate privileges from a regular user (low privilege) to root or administrator (high privilege). & By escalating privileges, attackers gain full control over the system, allowing them to perform high-privilege operations such as modifying configurations, installing software, or accessing sensitive data. & Directly exploits vulnerabilities to achieve privilege escalation, transitioning from low to high privilege. & Intruder \\
    \midrule[0.8pt]
    R2L & Attackers remotely gain access to local system permissions through network attacks, typically by stealing credentials or exploiting vulnerabilities. & Obtains local user access from a remote position, enabling further malicious activities within the system. & Gains local access through remote vulnerabilities or credential theft. & Intruder \\
    \bottomrule[0.8pt]
  \end{tabular}
  \label{A}
\end{table*}

Attackers employ a variety of methods to gain legitimate access to user accounts, primarily through system intrusions or by exploiting system vulnerabilities. These intrusions, particularly those utilizing User to Root (U2R) and Remote to Local (R2L) attack vectors, aim to acquire legitimate user permissions. U2R attacks typically commence with the attacker exploiting vulnerabilities in ordinary user accounts on the target system to gain root or administrative privileges. These attacks leverage flaws in the operating system, applications, or scripts that grant elevated privileges, thus allowing unauthorized access to administrative controls. In contrast, R2L attacks involve attackers sending packets from a remote location to exploit vulnerabilities in a system where they do not have an account, achieving local user-level access. This access enables attackers to reach confidential data. Following network-based intrusions, attackers often remain covert within the system, masquerading as legitimate users \cite{r1_6yuan2021deep}. These masqueraders typically exhibit two distinct behavioral patterns: one involves rapidly damaging the system upon gaining access, while the other entails creating backdoors to stay hidden within the system, posing as legitimate users and waiting for an opportune moment to act maliciously \cite{yadav2015technical,r1_5aldweesh2020deep,sommer2010outside}. To further elaborate on the related intrusion methods and the threats they pose, We have summarized the complex strategies employed by intruders as they transition from external to insider domains, as detailed in Table \ref{A}.

As the attack surface continues to expand, encompassing the advanced intrusion methods outlined in Table \ref{A}, current technologies are increasingly inadequate to address the evolving complexities and breadth of these sophisticated attack vectors. The underlying reasons can be traced back to the datasets. Existing insider threat datasets primarily focus on the privileges of legitimate users\cite{bin2022insider}. However, they lacked the more complex and evolving threats activity data posed by external attackers who gain legitimate user access through advanced exploitation techniques. This oversight creates a significant gap, as the boundary between external and insider threats becomes increasingly blurred due to the rising sophistication of attacks that originate externally but manifest as insider threats once legitimate access is compromised. Specifically, current insider threat datasets often fail to represent attack vectors involving Probe, U2R, and R2L, as highlighted in Table \ref{A} (Notably, in Table \ref{A}, the classification labels 'Potential Intruder' and 'Intruder' are used to represent 'users who may transition from being external threats to insider threats' and 'users who have already transitioned from being external threats to insider threats,' respectively). The pathways that allow attackers to transition from external to insider positions—such as remote access exploitation or privilege escalation—are critical for understanding the full impact of these threats. By neglecting to include these transition points, existing datasets do not provide the activity data of how attackers leverage external vulnerabilities to achieve internal control, ultimately leaving substantial gaps in detection and prevention strategies.

The lack of comprehensive data on how external threats evolve into insider threats severely limits the practical application of these datasets in enterprise security. In real-world environments, the ability to anticipate and recognize the fluid nature of threats that traverse external and internal boundaries is crucial. Without addressing these gaps, current insider threat datasets fall short of enabling security solutions that truly reflect the complex and interconnected nature of modern cyber threats.

To ensure the security of sensitive data and the continuity of business operations, institutions often implement Identity and Access Management (IAM) \cite{r1_7pal2019limitations} and Privileged Access Management (PAM) \cite{r1_8tep2015taxonomy} systems. These systems provide unified management of legitimate personnel access, ensuring the secure access to necessary resources while preventing unauthorized entries. A key component of IAM is UEBA \cite{r1_9diop2021high}, which assists in detecting anomalous behaviors among legitimate users to prevent damage to system assets. However, current UEBA technologies often struggle to differentiate between masqueraded and legitimate user behaviors, making masquerade attacks particularly challenging to detect. The evolution of these threats necessitates a detection strategy that not only identifies intrusions but also tracks the transition of these threats from external to internal. An effective detection and response system is crucial not only for quickly recognizing these changes, but also for mitigating risks before substantial damage occurs. This underscores the urgent need for continuous monitoring and advanced analysis of threats. Some practical applications of machine learning techniques \cite{liu2017singular,liu2020experiments} and recent developments in distributed and collaborative systems have introduced new challenges and opportunities for cybersecurity, particularly in areas such as cloud robotics and federated learning \cite{liu2022elasticros,liu2023roboec2,yan2021fedcm,zhang2022authros,liu2024edgeloc,zheng2022applications}. Such systems are essential for maintaining security and are also critical for enhancing the resilience of organizational IT ecosystems against increasingly dynamic and covert cyber threats. Numerous threat detection methodologies are currently employed, yet the effectiveness of these existing solutions is undermined by several critical issues:

\begin{enumerate}
    \item Existing methodologies face significant challenges in accurately distinguishing between normal user behaviors and those that are intentionally masqueraded \cite{r17_2khanna2021using,meng2018deep,sharma2020user,yuan2021deep}. This difficulty arises from the complexity involved in analyzing diverse behavioral patterns, including contextual, sequential, and temporal behavior patterns.
    \item The current research landscape has largely neglected the potential insider threats posed by Privilege Escalation Attacks and Remote Exploit Attacks \cite{chen2014study}. Existing methodologies for insider threat detection predominantly focus on monitoring authorized user entities, thereby overlooking the significant risks associated with unauthorized entities that may execute external attacks to usurp legitimate access rights \cite{verizon2008data,r17_1yuan2021deep}.
    \item Existing methods are difficult to detect rare types of attacks. In real network environments, there are often characteristics of diversified attack methods and attack characteristics that are not easy to identify, therefore these methods tend to have a high false alarm rate and cannot be put into practical production applications \cite{r11li2021sustainable,r7_2aburomman2016novel}.
    \item Amid the growing complexity of attacks that blur the lines between insider and external threats, current insider threat datasets exhibit significant limitations, particularly in terms of their breadth and depth \cite{bin2022insider,li2020deepfed}. Specifically, they overlook the critical pathways through which intruders and potential intruders transition from external probing and escalation to becoming full-fledged insider threats, as detailed in Table \ref{A}. While existing insider threat datasets provide foundational data support for threat detection, they fall short in terms of depth and comprehensiveness. Advanced Persistent Threats often involve attackers gaining access and remaining dormant for extended periods, awaiting the most opportune moment to act. This behavior is often underrepresented in existing datasets, which tend to focus more on immediate attack activities rather than long-term threats. This inadequacy hampers the ability of insider threat detection technologies to keep pace with the evolving complexity and expanding scope of these sophisticated attack vectors \cite{yadav2015technical,sommer2010outside}.
\end{enumerate}

To address these issues, we proposed the IDU-detector to enhance the detection capabilities for insider threat. We propose the DenseAttDNN Classifier, which includes dense connection and attention mechanism, termed DenseAttDNN. Within the classifier, dense connection promotes feature reuse, while the attention mechanism enables the model to focus on more crucial feature information. Therefore, this framework improves the model's detection capability, particularly for rare class attacks. In summary, our contributions can be listed as follows:

\begin{enumerate}
    \item We have proposed a novel framework aimed at detecting both insider and external security threats, tracking their transition from external to internal sources, and possessing specific capabilities for identifying intruder and potential intruder. The effectiveness of our approach has been validated using the datasets we proposed in the cybersecurity domain.
    \item Our work effectively detects privilege escalation attacks and remote exploit attacks, overcoming the limitations of traditional insider threat detection techniques that are constrained to monitoring legitimate users. Furthermore, it addresses the challenges posed by unauthorized attackers exploiting system vulnerabilities to escalate privileges. By doing so, our approach provides a more comprehensive and precise defense for system security, filling gaps in existing technologies and significantly enhancing the detection and mitigation of complex threats.
    \item We evaluate our model on the synergistic datasets. The experimental results show that our method has excellent classification performance, a faster response and a better performance compared with the existing methods. Also, it successfully balances accuracy, time, and computational cost and addresses the gap where the traditional approaches only seek to optimize or make the best use of resources without considering the characteristics of the real-world detection requirements.
    \item To enhance the capabilities of insider threat detection technologies in addressing the evolving complexity and expanding scope of sophisticated attack vectors, we provided datasets that encompass a broader range of threat scenarios. The integration of such datasets will not only more accurately capture the dynamic evolution of current threat landscapes but also aid organizations in building more robust defensive frameworks. Therefore, we propose KDD-UEBA, NSL-UEBA, CIC-UEBA, and KDD-UNBLogs as benchmark datasets for insider threat detection models. These datasets are designed to effectively support deep learning models in accurately identifying insiders, potential insider threats, and intrusion activities. For further details on these datasets. More details are shown in the Table XIV (Table XIV in Appendix B). 
\end{enumerate} 

\section{Related work}
Threat detection in cybersecurity can generally be classified into two categories: insider and external threat detection. This classification depends on whether the potential threat actor has legitimate access rights. External threat detection, also known as intrusion detection, refers to identifying attackers who gain unauthorized access to the system. Conversely, insider threats detection are typically used to detect user's behavior. 
\subsection{Intrusion Detection}
Mainstream intrusion detection techniques are AI-based approaches that categorize intrusion detection into single model-based and ensemble learning-based detection approaches based on the number of classifiers.
\subsubsection{Intrusion Detection Approaches Based on Single Model}
Single model-based intrusion detection means that IDS relies on a single algorithm or model to detect anomalies or attacks, and identifies suspicious and malicious intrusions by learning from anomalous or normal network connections. In early detection models, Jaiganesh et al. \cite{r1jaiganesh2013analysis} first proposed dynamic intrusion detection using Backpropagation Neural Network (BPN). Wei et al. \cite{r2wei2021deephunter} considered the threat detection task as a graph pattern matching problem and constructed a GNN-based graph pattern matching model using Graph Neural Network (GNN) for estimating the matching scores between the originating graphs and given query graphs, called DeepHunter. It's an end-to-end intrusion threat detection, which is achieved by using the attention mechanism to assign weights of different attributes of the attribute embedding network. However, intrusion behaviors are usually a string of consecutive behaviors in the time-series, and most models do not have the ability of time-series learning, which results in the loss of feature information of anomalous behaviors at the macro level. To solve this problem, Guo et al. \cite{r3guo2021network} proposed a multi-head self-attention mechanism (Transformer) detection method based on the reduced dimensional features, combined with the transformer pair having the ability to model both long-term and short-term time-series features, which effectively improves the detection efficiency. Jin et al. \cite{r4jin2020swiftids} improved on the real-time detection by proposing a model called SwiftIDS model. This model uses Light Gradient Enhancement Machine (LightGBM) and parallel detection mechanism to effectively deal with massive data in high-speed networks. Imrana et al. \cite{r5imrana2021bidirectional} proposed an intrusion model based on Bidirectional Long Short Term Memory (BiDLSTM), which acquires input data and its reverse copy on two subnetworks, adding more meaningful information to the network.

Models based on a single algorithm can streamline computations, enhance processing speeds \cite{r5_1roshan2018adaptive,r5_2jeong2017mean}, and optimize the utilization of computing and storage resources. However, they often encounter several challenges. Scalability is a primary issue, as such models may not handle large datasets efficiently, leading to increased computational demands. Additionally, they are susceptible to overfitting, where the model excessively learns from the noise in the training data, subsequently impairing its performance on new, unseen data. Generalization is another concern; these models might not perform well with data or scenarios that differ from those encountered during training. Moreover, these models often lack the flexibility required to process various types of data inputs or perform different tasks without substantial modifications. Lastly, its performance is heavily dependent on the quality of input data; issues such as noise and outliers can significantly affect its effectiveness.
\subsubsection{Intrusion Detection Approaches Based on Ensemble Learning}

Currently, the most popular models in intrusion detection are ensemble learning models. Intrusion detection based on ensemble learning refers to predicting behavior by combining the results of multiple models. In the context of IDS, it implies the use of multiple algorithms and learning modules to make a joint decision to determine the presence of intrusion behavior. Ensemble learning mainly includes several basic techniques: voting, bagging, boosting, and stacking \cite{r7_1aburomman2017survey}. These methods effectively reduce the model's bias and variance, are suitable for various data distributions, and significantly improve the model's ability to handle complex data and the reliability of predictions. Although there are many ensemble methods available, finding the right ensemble configuration for a specific dataset remains a challenging task. Aburomman et al. \cite{r7_2aburomman2016novel} proposed a novel approach to constructing ensembles, utilizing weights generated by particle swarm optimization to create more accurate classifier ensembles.

Teng et al. \cite{r6teng2017svm} constructed an adaptive collaborative intrusion detection framework that outperforms a single type of SVM approach by using SVM and DT methods. For the hybrid intrusion detection system scheme with SVM support. Jabbar et al. \cite{r7jabbar2017cluster} viewed intrusion detection as an access behavioral clustering problem, by combining the KNN with ADTree to construct a clustering-based IDS integrated classifier to achieve detecting behavioral intent. In the real-world scenarios, Moustafa et al. \cite{r8moustafa2018ensemble} in order to mitigate botnet attacks encountered in the IoT, developed an AdaBoost integrated learning approach to assess the impact of these malicious attack features and detect intrusion events effectively. For the case where the network data traffic in the source data is too noisy and the feature library is difficult to be updated; the data volume of host log files is limited and the intrinsic information needs to be fully mined. Wang et al. \cite{r9wang2021intrusion} proposed an integrated deep intrusion detection model based on SDAE-ELM by integrating the SDAE-ELM and DBN-Softmax frameworks to detect intrusions on different datasets (NIDS and HIDS) for separate detection, and obtains new improvements in training speed and accuracy. The above methods still have poor detection performance for network traffic with multiple features and unbalanced classes. Zhou et al. \cite{r10zhou2020building} constructed a new intrusion detection framework based on feature selection and ensemble learning techniques, choosing the heuristic CFS-BA dimensionality reduction algorithm in combination with the random forest to learn the features and construct a decision tree, and finally conducting joint voting for attack identification. It shows good performance on the NSL-KDD dataset. Li et al. \cite{r11li2021sustainable} proposed a model based on sustainable ensemble learning, which obtains the weights of different attack types through multiple classifiers, and carries out the accumulation and transfer of feature knowledge, which improves the accuracy of model detection. Talukder et al. \cite{r12talukder2023dependable} proposed a hybrid detection model combining SMOTE and XGBoost, which achieved 100\% accuracy in both binary classification tasks on the CIC-MalMem-2022 dataset by combining RF and Artificial Neural Network (ANN) models. However, the model's ability to deal with unbalanced datasets mainly relies on SMOTE, and the recognition ability is insufficient when the experimental data is not expanded in proportion to the dataset, and remains weak in the face of novel attack patterns. In the context of Telematics security, Yang et al. \cite{r13yang2021mth} proposed a multi-layer hybrid intrusion detection system, MTH-IDS, which combines DT, RF, Extra Tree (ET), and Extreme Gradient Enhancement (XGBoost) as classifiers to detect known attacks, and K-means to detect zero-day (unknown) attacks. To address the challenge of existing solutions struggling to adapt to the diversity of attacks and dynamic temporal variations. Li et al. \cite{r13_1li2021sustainable} proposed a sustainable Ensemble Model, during the model training phase, we constructed a multi-class regression model by using the probability outputs and classification confidence of individual classifiers as training data, enabling the ensemble learning to adapt to different attacks. Additionally, in the update phase, an iterative updating method was proposed, incorporating the parameters and decision outcomes of historical models into the training process of the new ensemble model, achieving incremental learning.

Although these studies have made interesting contributions to improving intrusion detection technology, most of them overlook the fact that in the real world, intrusion activities are rare class attacks, making the detection of rare classes more meaningful for practical applications. Since the above approaches did not consider the transfer and reuse of features, the classifiers failed to adequately learn about rare class attacks, thus reducing detection accuracy. To address these issues, this paper resamples rare class samples and incorporates dense connections into the classifiers, enabling the classifiers to better learn the rare class features while achieving high overall accuracy. This enhances the accumulation and transfer of feature knowledge, further improving the stability and certainty of the model.
\begin{table*}[bp]
\linespread{1.5}
\centering
\caption{Supporting features of our framework compared to the available literature}
\fontsize{10}{8}\selectfont
\renewcommand{\arraystretch}{1} 
\resizebox{\textwidth}{!}{%
\begin{tabular}{p{3cm}p{1.5cm}p{1.5cm}p{2.5cm}p{2.5cm}p{2.5cm}p{2.5cm}p{1cm}}
\toprule[0.8pt]
\textbf{Previous Work} & \textbf{Efficiency} & \textbf{Robustness} & \textbf{Imbalanced Data Handling Capability} & \textbf{Potential Intruder Detection} & \textbf{Privilege Escalation Attacks(U2R) Detection} &\textbf{ Remote Exploit Attacks(R2L) Detection}   \\
\midrule[0.8pt]
Gavai et al. \cite{r14gavai2015detecting}   & \checkmark & -- & -- & -- & -- & -- \\
Kim et al. \cite{r15kim2019insider}         & \checkmark & \checkmark & -- & -- & -- & -- \\
Zhang et al. \cite{r16zhang2021detecting}   & -- & -- & \checkmark & -- & -- & -- \\
He et al. \cite{r17he2021insider}           & -- & \checkmark & \checkmark & -- & -- & -- \\
Khanna et al. \cite{r17_2khanna2021using}   & \checkmark & \checkmark & -- & -- & -- & -- \\
\textbf{Ours }                                      & \checkmark & \checkmark &\checkmark & \checkmark & \checkmark & \checkmark \\
\bottomrule[0.8pt]
\end{tabular}%
}
\label{tab:issues}%
\end{table*}

\subsection{Insider Threat Detection Approaches}
Insider threats are cybersecurity threats initiated by or involving authorized users within an organization who intentionally or unintentionally misuse their legitimate access privileges or whose accounts have been hijacked by cybercriminals.IBM Corp. categorizes insider threats into three categories \cite{r17_1yuan2021deep}: misuser attacks, masquerade attacks, and malicious user attacks.

Earlier detection approaches used more machine learning methods to identify anomalous states of manually extracted user behavioral features. Gavai et al. \cite{r14gavai2015detecting} detected insider threats in organizations by analysing the social and online activity data of the employees in the organization. Kim et al. \cite{r15kim2019insider} combined Gaussian density estimation (Gauss), window density estimation (Parzen), principal component analysis (PCA) and K-mean clustering to construct classification algorithms for the detection framework. Zhang et al. \cite{r16zhang2021detecting} proposed a behavioral log detection method based on ensemble learning and self-supervised learning, which employs the entity embedding method of TF-IDF, and performs a self-supervised classification task on the detector inputs, which directly distinguishes between the normal and malicious behaviors of a particular user. 

In subsequent research work, deep learning feed-forward networks, networks such as CNN, RNN, GNN, and some other representation learning techniques have been used to automatically extract user behavioral features with good results in fitting analysis. Many methods focus on detecting anomalous behavior, but these approaches ignore the connectivity information between entities, which can result in the loss of important information. He et al. \cite{r17he2021insider} proposed an approach based on historical user behavioral information and attention mechanisms. The fine-grained detection of anomalous user behavior is achieved by learning the differences between different behavioral patterns through ABUHB (attention) and predictive modelling of user behavior using bidirectional LSTM. To further enhance the predictive performance of the model, Khanna et al. \cite{r17_2khanna2021using} developed a new method to extract robust features and enhance attack vectors. They utilize high-quality color image encoding to represent user behavior, avoiding the limitations of traditional grayscale image encoding and improving classification capabilities.

In practical insider threat detection, many models fail to analyze user activity patterns from multiple perspectives, resulting in poor stability when identifying rare threat categories and high false alarm rates. This single-dimensional analysis approach limits the models' comprehensive understanding of complex threat scenarios, thereby affecting the accuracy and reliability of detection. To enhance the effectiveness and precision of detection systems, there is an urgent need to adopt more comprehensive and refined behavior analysis methods. Moreover, current insider threat detection technologies primarily focus on analyzing behavioral features, often overlooking complex threat scenarios where users gain legitimate access through network or host intrusions. This limitation not only results in persistently high false alarm rates and significantly reduced detection accuracy but also hampers the ability of these technologies to address the evolving sophistication of advanced attack methods. Consequently, the overall effectiveness and reliability of these models in real-world applications are compromised. To address these challenges, there is an urgent need to develop more comprehensive threat datasets that can fill the gaps in current technologies, thereby enhancing the threat detection capabilities of machine learning models in diverse threat environments.

As highlighted in the Introduction section and detailed in Table \ref{A}, network-originated attacks can significantly impact insider security, which underscores the need to incorporate a broader scope of threats beyond typical insider scenarios. Consequently, when addressing insider threats, it is crucial to also consider threats from various parts of the network that could compromise legitimate access privileges. As shown in Table \ref{tab:issues}, our study aims to bridge the gaps present in existing research by addressing security issues such as U2R and R2L attacks, which are often overlooked. To achieve this, we employ novel techniques that enhance the detection framework. Our proposed framework integrates IDS with UEBA, thereby creating a more comprehensive approach to insider threat detection.

\section{Methodology}
In this section, we outline our workflow and the specific organizational architecture of our model, providing detailed descriptions of each component.

\begin{figure*}[!htb] 
\centering
\includegraphics[width=\textwidth]{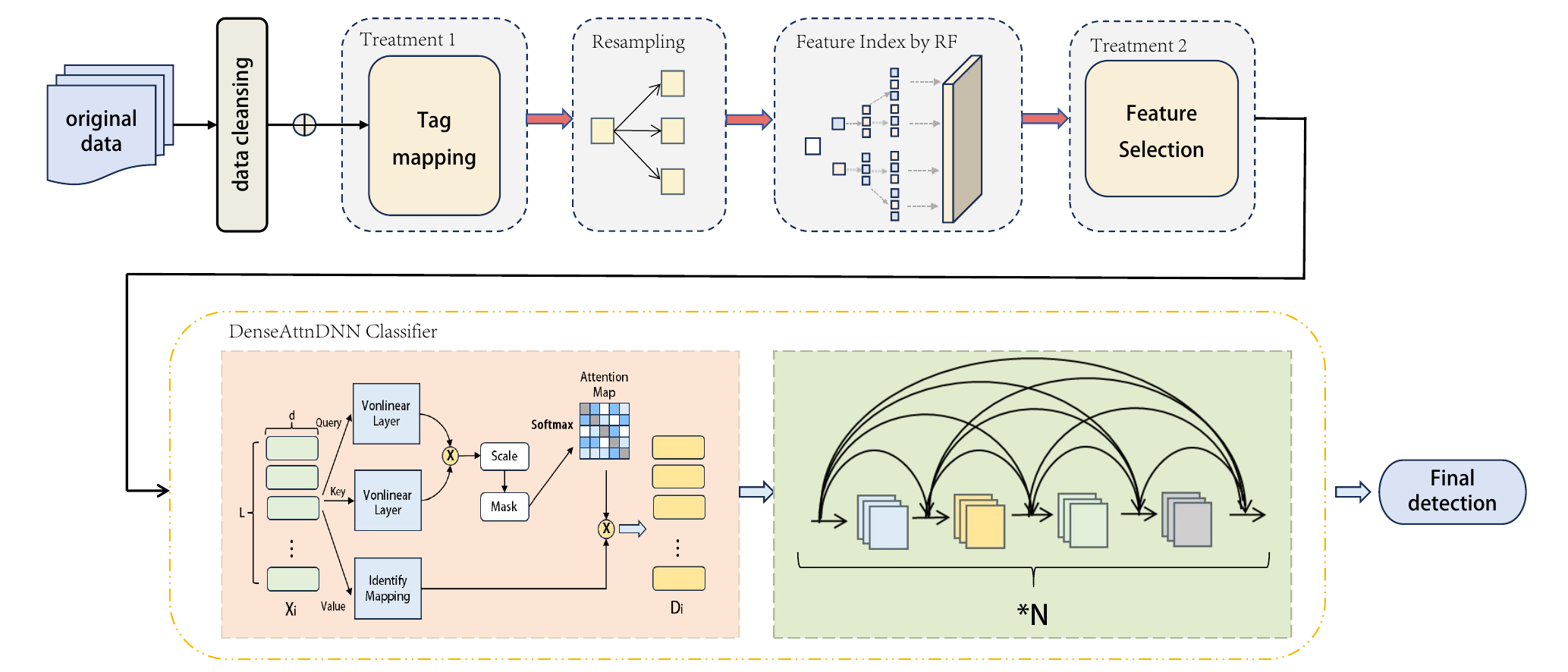} 
\caption{Pipeline of IDU-Detector}
\label{fig1}
\end{figure*}

\subsection{Overview}
Our approach takes network traffic and user behavior records as input and outputs prediction results based on various attack vector features. As illustrated in Fig. \ref{fig1}, the pipeline of our method consists of four main components: Data Preprocessing, Resampling, Feature Selection, and the classifier.

\textbf{Data Preprocessing.}
This section involves data synthesis, processing, and encoding. The encoding part includes using ColumnTransformer for data feature encoding and applying one-hot encoding to label, making the data compliant with the input specifications of the IDU-Detector. Also, we map the input tag to the corresponding attack class, and the mapping rules are as Table \ref{tab1:KDDCup99 Dataset Tags Mapping}, \ref{tab2:KDDCup99 Dataset Tags Mapping}, \ref{tab3:CICIDS2017 Dataset Tags Mapping} and Table XIV (Table XIV in Appendix B). As detailed results and supplementary figures are provided in the appendix, they have been uploaded to Google Drive for easier access. The appendix can be accessed via the following link: \url{https://drive.google.com/file/d/1RiTJ-PY_LR72mCC1x7kxjxQlF5i-Slzt/view?usp=sharing}.

\textbf{Resampling.}
It involves resampling of rare class attack samples to produce effective attack instances while maintaining a distribution ratio close to that of the original data. This enhances the model's predictive ability regarding rare class attacks.

\textbf{Feature Selection.}
Different datasets vary in input dimensions and often contain multiple irrelevant features. To reduce noise in the data, facilitate the model's ability to capture principal features, minimize the risk of overfitting, and enhance the model's generalization ability, we employ 
Random Forest for dimensionality reduction in this section.


\textbf{DenseAttDNN.}
We employed the proposed DenseAttDNN for classification. The methodology is elaborated in detail in the following subsections.


\begin{table*}[ht]
  \centering
  \caption{KDDCup99 Dataset Tags Mapping}
  \label{tab1:KDDCup99 Dataset Tags Mapping}
  \fontsize{10}{12}\selectfont
  \begin{tabularx}{\textwidth}{@{}>{\hspace{5mm}}p{0.1\textwidth}p{0.3\textwidth}X@{}} 
    \toprule[0.8pt]
    Class & Description & Related Tags \\
    \midrule[0.8pt]
    Probe & Probing attack & ipsweep; nmap; portsweep \\
    Dos & Denial of service attacks & smurf; neptune; back; pod; land; teardrop \\
    U2R & User to root permission attacks & buffer\_overflow; rootkit; loadmodule; perl \\
    R2L & Remote to local permission attacks & ftp\_write; imap; multihop; phf; spy; warezclient; guess\_passwd; warezmaster \\
    Benign & Normal & BENIGN \\
    \bottomrule[0.8pt]
    \label{tab1}
  \end{tabularx}
\end{table*}

\renewcommand{\arraystretch}{1} 
\begin{table*}[]
  \centering
  \caption{NSL-KDD Dataset Tags Mapping}
  \fontsize{10}{12}\selectfont
  \label{tab2:KDDCup99 Dataset Tags Mapping}
  \begin{tabularx}{\textwidth}{@{}>{\hspace{5mm}}p{0.1\textwidth}p{0.3\textwidth}X@{}} 
    \toprule[0.8pt]
    Class & Description & Related Tags \\
    \midrule[0.8pt]
    Probe & Probing attack & satan; ipsweep; nmap; portsweep; mscan; saint \\
    Dos & Denial of service attacks & \makecell[lt]{neptune; worm; smurf; back; pod; land; apache2; mailbomb;\\ processtable; teardrop; snmpgetattack; httptunnel; sqlattack; udpstorm} \\
    U2R & User to root permission attacks & buffer\_overflow; loadmodule; rootkit; perl; xterm; ps \\
    R2L & Remote to local permission attacks & \makecell[lt]{guess\_passwd; ftp\_write; imap; phf; multihop; warezmaster;\\ spy; named; sendmail; snmpguess; xlock; xsnoop; 
 warezclient} \\
    Benign & Normal & BENIGN \\
    \bottomrule[0.8pt]
  \end{tabularx}
  \label{tab2}
\end{table*}

\renewcommand{\arraystretch}{1} 
\begin{table*}[]
  \centering
  \caption{CICIDS2017 Dataset Tags Mapping}
  \fontsize{10}{12}\selectfont
  \label{tab3:CICIDS2017 Dataset Tags Mapping}
  \begin{tabularx}{\textwidth}{@{}>{\hspace{5mm}}p{0.1\textwidth}p{0.3\textwidth}X@{}} 
    \toprule[0.8pt]
    Class & Description & Related Tags \\
    \midrule[0.8pt]
    Probe & Probing attack & PortScan \\
    Dos & Denial of service attacks & DoS slowloris; DoS Slowhttptest; DoS Hulk; DoS GoldenEye; DDoS \\
    U2R & User to root permission attacks & Heartbleed \\
    R2L & Remote to local permission attacks & \makecell[lt]{Infiltration; FTP-Patator; SSH-Patator; Web Attack – Brute Force;\\ Web Attack – XSS; Web Attack – Sql Injection} \\
    Benign & Normal & BENIGN \\
    \bottomrule[0.8pt]
  \end{tabularx}
  \label{tab3}
\end{table*}

\subsection{Detailed}
DenseAttDNN forms the core of the IDU-Detector framework and is utilized for classifying types of attacks. DenseAttDNN is developed by incorporating dense connections and attention mechanism into the fundamental DNN classifier architecture. The basic DNN framework consists of an input layer, several hidden layers, and an output layer. The hidden layers are primarily fully connected, where each neuron is connected to every neuron in the previous layer. Activation functions like ReLU are employed between these layers to introduce non-linearity, enabling the network to learn complex patterns. Furthermore, the DNN includes regularization layers such as BatchNorm1d and Dropout; BatchNorm1d is used for batch normalization, while Dropout helps to accelerate the training process and reduce overfitting.

\begin{figure*}[bp]
\centering
\includegraphics[width=0.6\textwidth]{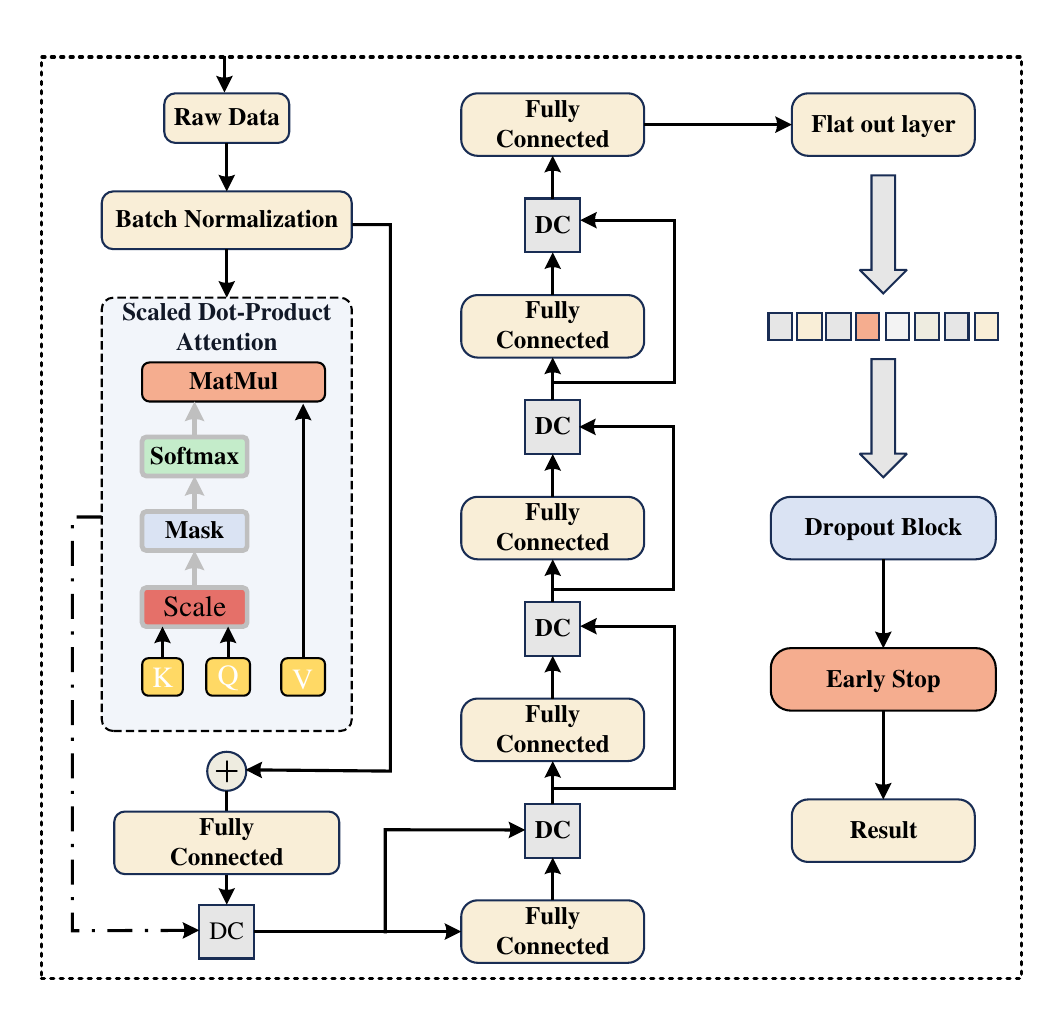}
\caption{Architecture of DenseAttnDNN Classifier}
\label{fig2}
\end{figure*}

We transform the threat detection problem into a classification problem on the acquired detection features with corresponding correctly labeled results $\{D_n\}$ for a given input $N$ sets of unclassified data $\{X_n\}_{1 \leq n \leq N}$. Our goal is to learn a nonlinear mapping $F$ that can be mapped to ground truth by the abstract parameter $\phi$ on the input data, which holds in the condition $D_n (X_n) = F(\phi, X_n)$. The network architecture is shown in Fig. \ref{fig2}. Since the initial cleaned data still suffers from excessive variance, the input training data is scaled and shifted by one-dimensional batch normalization, as a rule, to improve the training speed and stability of the network.
\begin{equation}
\label{e1}
\hat{x}_i = \frac{x_i - \mu}{\sqrt{\sigma^2 + \epsilon}}
\end{equation}
We first calculate the mean $\mu$ and variance $\sigma^2$ for each feature data in the small batch, while $\epsilon$ denotes here a non-zero minimal constant that prevents division by zero in \ref{e1}.

\begin{equation}
\label{e2}
y_i = \gamma\hat{x}_i + \beta
\end{equation}

Here $\gamma$ and $\beta$ are a scaling factor that can be automatically learned through network training and an offset that can also be learned to adjust the scale of the feature data in the batch normalization process to better manipulate the original feature data in \ref{e2}. However, it is similar to image convolution, the greater variation reflects the higher concentration of features in the data, and it is more important to give more attention to this part of the data to improve the concentration of features in this part of the feature. To allow the model to focus on more relevant information and improve processing efficiency and accuracy, we introduced the Scaled Dot-Product attention mechanism, a way to calculate the weight of attention. The attention mechanism is executed by aggregating a collection of queries into a singular matrix denoted as $Q$. Correspondingly, key and value sets are consolidated into their respective matrices, labeled $K$ and $V$. The outcome is an output matrix derived as \ref{e3}.
\begin{equation}
\label{e3}
\text{Attention}(Q, K, V) = \text{softmax}\left(\frac{QK^T}{\sqrt{d_k}}\right)V
\end{equation}

The two primary forms of attention functions, additive, and scaled dot-product, share a level of theoretical complexity. However, the scaled dot-product variant has a practical advantage in terms of speed and memory efficiency due to the use of refined matrix multiplication algorithms. When $d_k$ is minimal, both attention functions deliver comparable results. Nevertheless, as $d_k$ increases, the additive model tends to surpass the non-scaled dot-product approach. This is hypothesized to occur because the magnitude of the dot products intensifies with larger $d_k$, which can lead the softmax function into a zone of minimal gradient values. To mitigate this, the dot products are adjusted by the factor $\frac{1}{\sqrt{d_k}}$.

The simplest way to use our perceptual features is to link the features processed by the self-attention mechanism to the normalized result $\hat{x}_i$, however, this does not account well for the classification problem in a multi-target setting, where the current matrix information remains relatively shallow and lacks explicitly pointing identifiers for multi-target judgments. We then propose a new network structure to better process the features of the data by utilizing multiple fully connected layers. In order to fully exploit and utilize the complex patterns in the data, for this purpose, we first define the rich feature extraction matrix as \ref{e4}.

\begin{equation}
\label{e4}
R_i = y_i + A_i
\end{equation}

In practice, we use $L = 7$ consecutive fully-connected layers of different scales $\omega_j = (512,512,256,128,64,32,16)$ to perform consecutive operations on the feature-rich extraction matrix, $R_i \in K_n \times K_m $, $ \omega $ which is the matrix consisting of a length of $K_n$, a height of $K_m$, and the number of channels $\omega$. In this paper, with the fully-connected layer, we can perform a cross-channel combination of contextual information without changing the size of the matrix to further extract information. Experiments confirm that the effectiveness of the mapping will be degraded without this step. We compute these features as \ref{e5}, \ref{e6}.

\begin{equation}
\label{e5}
y = xW^T + b
\end{equation}

\begin{equation}
\label{e6}
R_L = [R_{L-1}|\text{ReLU}(R_L \cdot \omega^T_j + b_L)]
\end{equation}
where $[|]$ denotes a channel cascade operation that channel splices the outputs of the layer with the outputs of the fully-connected network in this layer. It is here that this cascading fully connected layer structure allows the network to progressively abstract and refine the input features, with each layer building on the previous one to construct more complex feature representations, abstracting more and more advanced features. Each batch of samples, represented as a uniquely specific one-dimensional array, converts the flattened feature vectors into a final output that is used to evaluate the final classification of the sample categories, represented as \ref{e7}.
\begin{equation}
\label{e7}
\text{flatten}(R_{L=7}) = [R_{L=7}^{1,1}, R_{L=7}^{1,2}, \ldots, R_{L=7}^{n,m}]
\end{equation}

\begin{figure}[t]
\centering
\includegraphics[width=0.48\textwidth]{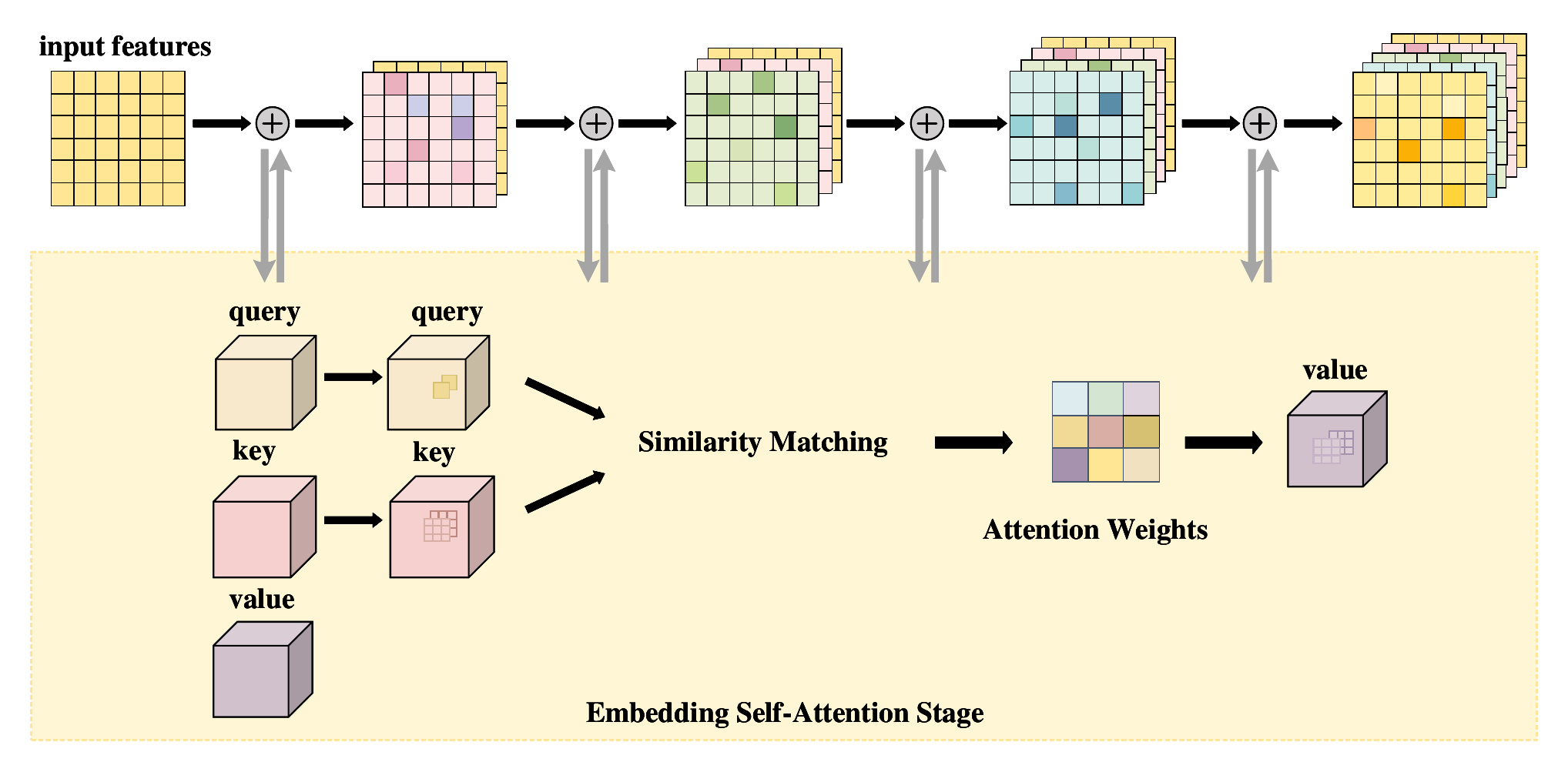}
\caption{Attention Map-Driven Feature Flow in Dense Connections}
\label{fig3}
\end{figure}
In order to enhance the flow of information between the layers inside the neural network, we borrowed the architecture of DenseNet and introduced a new Connection mode called Dense Connection, which enables each layer to access all the characteristic information provided by all the previous layers. This approach establishes a direct link from any given layer to each subsequent layer, better preserving the original characteristics of the data. This method establishes direct links from any given layer to every subsequent layer. Depicted in the Fig. \ref{fig3} and Fig. 5 (Fig. 5 in Appendix A), this schematic of DenseNet reveals how each layer $\ell$ integrates feature-maps from all preceding layers $x_0,x_1,\ldots,x_{\ell-1}$ as its input through concatenation:

\begin{equation}
\label{e8}
x_{\ell} = H_{\ell}([x_0, x_1, \ldots, x_{\ell-1}])
\end{equation}

In \ref{e8}, $[...]$ denotes the concatenation of the feature-maps generated by layers 0 through $\ell - 1$. This architecture is termed Dense Network due to its dense inter-layer connections. To simplify, we merge the multiple inputs of $H_{\ell}$, as stated in \ref{e8} into a single tensor for implementation convenience. The composite function $H_{\ell}$ is defined by a sequence of three operations. These operations include batch normalization (BN), a rectified linear unit (ReLU), and a fully-connected layer, performed in this specific order.

In the DenseAttDNN classifier, each layer \( l \) receives as input the concatenated outputs of all preceding layers. As shown in \ref{ee1}, this dense connectivity ensures that the network captures a comprehensive range of features from different hierarchical levels. The output at layer \( l \) can be expressed as:
\begin{equation}
\mathbf{H}_l = \sigma\left(\mathbf{W}_l \cdot \left[\mathbf{Z}_0, \mathbf{H}_1, \dots, \mathbf{H}_{l-1}\right] + \mathbf{b}_l\right)
\label{ee1}
\end{equation}

where \( \mathbf{H}_l \) is the output of layer \( l \) after applying the non-linear activation \( \sigma(\cdot) \), \( \mathbf{W}_l \) and \( \mathbf{b}_l \) are the weight matrix and bias vector associated with layer \( l \). The dense connections enable effective gradient propagation through the network, avoiding vanishing gradients, which is crucial for training deep architectures. The gradient flow can be described by \ref{ee2}.

\begin{equation}
 \frac{\partial \mathcal{L}}{\partial \mathbf{W}_l} = \sum_{i=l}^{L} \frac{\partial \mathcal{L}}{\partial \mathbf{H}_i} \cdot \frac{\partial \mathbf{H}_i}{\partial \mathbf{H}_l} \cdot \frac{\partial \mathbf{H}_l}{\partial \mathbf{W}_l}   
\label{ee2}
\end{equation}

This sum across all subsequent layers \( i \) ensures that the learning signal from the loss \( \mathcal{L} \) is preserved, improving training dynamics and performance. The self-attention mechanism in DenseAttDNN allows each feature to dynamically adjust its importance based on the entire feature set. For each layer \( l \), the query, key, and value vectors are computed as \ref{ee3}.

\begin{equation}
\mathbf{Q}_l = \mathbf{W}_q \cdot \mathbf{Z}_l, \quad \mathbf{K}_l = \mathbf{W}_k \cdot \mathbf{Z}_l, \quad \mathbf{V}_l = \mathbf{W}_v \cdot \mathbf{Z}_l    
\label{ee3}
\end{equation}

The attention mechanism then computes the weighted sum of the values, with the weights determined by the similarity between queries and keys, as shown in \ref{ee4}.

\begin{equation}
\text{Attention}_l(\mathbf{Q}_l, \mathbf{K}_l, \mathbf{V}_l) = \text{softmax}\left(\frac{\mathbf{Q}_l \cdot \mathbf{K}_l^\top}{\sqrt{d_k}}\right) \cdot \mathbf{V}_l    
\label{ee4}
\end{equation}

where \( \mathbf{Q}_l \), \( \mathbf{K}_l \), and \( \mathbf{V}_l \) are the query, key, and value matrices for layer \( l \), The softmax function normalizes the attention weights, \( d_k \) is the dimensionality of the key vectors, used to scale the dot product. The attention output is combined with the dense output to maintain both the comprehensive feature usage and the dynamic importance of features. The updated output for the next layer can be expressed as formula \ref{ee5}. This combined output \( \mathbf{Z}_{l+1} \) is then used as input to the next layer, continuing the process. The final output of the DenseAttDNN model is constructed by aggregating the features from all layers \ref{ee6}.

\begin{equation}
\mathbf{Z}_{l+1} = \text{Attention}_l(\mathbf{Q}_l, \mathbf{K}_l, \mathbf{V}_l) + \mathbf{H}_l    
\label{ee5}
\end{equation}

\begin{equation}
\mathbf{Z}_{\text{final}} = \left[\mathbf{Z}_0, \mathbf{H}_1, \dots, \mathbf{Z}_L\right]    
\label{ee6}
\end{equation}

This aggregated feature vector is then passed through a fully connected layer, followed by a softmax function to produce the final classification probabilities \ref{ee7}.

\begin{equation}
\hat{y} = \text{softmax}\left(\mathbf{W}_{\text{final}} \cdot \mathbf{Z}_{\text{final}}\right)    
\label{ee7}
\end{equation}

This comprehensive and mathematical approach allows DenseAttDNN to excel in capturing intricate dependencies in data, leading to superior performance across a wide range of classification tasks. This comprehensive and mathematically approach allows DenseAttDNN performs well in capturing intricate dependencies in features. In summary, as illustrated in Fig. 2, the DenseAttDNN classifier integrates a self-attention module with a layer-by-layer refinement process within the fully connected layers. This architecture enables the extraction of all preceding layer information into a singular one-dimensional vector. Such a configuration aids in capturing and assimilating the feature information across different layers, thereby enhancing the model's ability to detect hidden threats in the original data. This approach ensures a comprehensive analysis of inter-layer dynamics, contributing significantly to the model's predictive accuracy and robustness.

\section{Experiment Setup}
In this section, we provide a detailed description of the experimental environment in which our study was conducted. Additionally, we introduce authoritative datasets currently utilized in the fields of intrusion detection and insider threat analysis, alongside the performance evaluation metrics employed to assess model efficacy. Furthermore, we identify and discuss the existing models selected for comparison, laying the groundwork for subsequent performance evaluation of our proposed model.
\subsection{Experiment Environment}
Our model training was conducted on the Google Colab platform. We trained on computing resources equipped with an NVIDIA A100 GPU with 83.0GB of RAM, ensuring ample memory space to process our datasets and models. Our model was implemented and tested in Python 3.10.12 and PyTorch 2.1.0.
\subsection{Comparative Approaches Selection}
For the purpose of intrusion detection comparison, our analysis includes a comparison with two leading-edge methodologies. These are predominantly advanced ensemble learning models, specifically the Sustainable Ensemble Learning (SEL)\cite{r11li2021sustainable}, DeepFed\cite{li2020deepfed} and the SVM-KNN-PSO (SKP)\cite{r7_2aburomman2016novel} frameworks. 

In the domain of insider threat detection, our comparison encompasses three distinct methodologies: the deep learning-based DualNet-CV\cite{r17_2khanna2021using} architecture and two traditional classification models, namely SVM and Logistic Regression (LR).

\subsection{Dataset}
In this study, we innovatively integrated multiple publicly available datasets to provide a dataset that better aligns with current threat trends. We strategically selected the classic intrusion detection datasets listed in Table XII and the specialized insider threat datasets in Table XIII for comprehensive integration and optimization, resulting in a unified benchmark dataset as presented in Table XIV. These datasets cover a diverse range of network traffic characteristics and behavior patterns, encompassing various threat types and complex scenarios. By providing rich and multidimensional data, these datasets significantly enhance the models' ability to learn from threat scenarios, thereby enabling them to exhibit greater adaptability and responsiveness when confronted with constantly evolving and highly dynamic insider threats.

It is worth noting that the characteristics presented in Table XIV (Table XIV in Appendix B), including the attack categories, relevance to the study, and advantages, are shared across all four proposed datasets (CIC-UEBA, NSL-UEBA, KDD-UEBA, and KDD-UNBLogs). This uniformity reflects the datasets' comprehensive design, which enables consistent performance across various cybersecurity applications. The datasets mentioned in this subsection are available in Appendix B for detailed reference.

\subsection{Metrics}
The evaluation of the model's performance across datasets utilizes key metrics such as Accuracy, Precision, Recall, Detection Rate (DR), False Alarm Rate (FAR), False Negative Rate (FNR), and F1-Score.

\subsection{Tag-Mapping}
For intrusion detection, we employ the label mapping methods shown in Tables \ref{tab1}, \ref{tab2}, and \ref{tab3}. 
For insider threats, we use the label mapping method outlined in Table XIII.
For synergistic detection, we apply the label mapping approach detailed in Table XIV (Table XIII and Table XIV in Appendix B).

\section{Experimental Analysis}
In this section, we conduct a theoretical analysis of the model's time complexity and evaluate its performance using the metrics in the Metrics section. To demonstrate that our experimental results are not coincidental, we repeated the experiments 10 times to ensure the model's high stability in practical applications.
\subsection{Theoretical Analysis of IDU-Detector}
In networks, especially those expecting the high throughput associated with 5G and 6G technologies, the efficiency of Threat Detection Systems (TDS) is crucial. Should a TDS fail to keep pace with data flow, it could impair network performance and become a bottleneck. Therefore, it is essential for TDS to maintain high detection efficiency, ensuring network security without compromising on throughput.

We have studied the training time complexity of existing state-of-the-art classifiers, and the results are presented in Table \ref{tab4}. In this table, \(n\) denotes the scale of the problem. It can be seen that those models can be divided into two categories according to the number of classifiers. These models are the mainstream classification models for their high detection efficiency. The first category includes DT, RF, and KNN, with their time complexities being \(O(n\log n)\) and \(O(n)\). Among these, KNN has the optimal time complexity, which is \(O(n)\), however, within the context of TDS, due to the scarcity of rare class samples, these samples may be far apart from each other. This implies that even samples within the same class are unlikely to be selected as nearest neighbors, thus increasing the risk of misclassification of rare class samples. Therefore, KNN is not suitable for TDS.

The other category comprises XGBoost and Gradient Boosting Decision Tree (GBDT), both of which share the same training time complexity \(O(n\log n)\). Our model belongs to the first category, and its time complexity is lower than that of any other model with which we have compared it. Although the interpretability of tree models is a significant advantage in the security domain, as it can aid in analyzing and explaining detected threats, IDS typically require real-time or near-real-time data processing to quickly detect and respond to potential security threats. Tree models, especially complex ensemble tree models, may be slow when processing large volumes of data, making them potentially unsuitable for high-speed data streams or real-time processing requirements. Therefore, XGBoost and GBDT may not be suitable for IDS.

{
\renewcommand{\arraystretch}{1.3} 
\fontsize{10}{12}\selectfont
\begin{table}[ht]
  \centering
  \caption{Model Time Complexity}
  \label{tab4:model-time-complexity}
  \begin{tabular}{lc}
    \hline
    \textbf{Models} & \textbf{Time Complexity} \\
    \hline
    RF & \(O(n\log n)\) \\
    KNN & \(O(n)\) \\
    DT & \(O(n\log n)\) \\
    XGBoost & \(O(n\log n)\) \\
    GBDT & \(O(n\log n)\) \\
    \textbf{ours} & \textbf{\(O(n)\)} \\
    \hline
  \end{tabular}
  \label{tab4}
\end{table}
}

\subsection{Experiment Result Analysis}
In this section, we compare classifiers for intrusion detection and insider threats and analyze our experimental results.

\subsubsection{Analysis of Intrusion Detection Effectiveness}
In our intrusion detection experiment, we evaluated several classifiers of our model on the CIC-UEBA and KDD-UEBA datasets. The results of each detection method are shown in Tables \ref{table5} and \ref{KDD-UEBA1}. Our approach demonstrated strong performance across four key metrics: Accuracy, Precision, Recall, and F1-Score. Notably, our method not only accurately identified common attack types, such as DoS and Probe, but also excelled in detecting more complex and elusive attacks, such as R2L and U2R.

In the SKP model, the Particle Swarm Optimization (PSO) and Local Update Strategy (LUS) methods used to optimize classifier weights did not allocate specific weights to different types of attacks. Instead, a uniform weight distribution was applied to all attack types. This approach overlooks the potential differences among various attack types, leading to suboptimal performance, particularly in handling rare attack types. Furthermore, the study did not mention the use of specific techniques to enhance the transfer and reuse of features, which may result in the consideration of excessive unnecessary redundant features during the learning process, thus impacting the model's learning efficiency and performance. The lack of appropriate weighting and strategies for reusing critical features may impair the model's ability to distinguish between normal behavior and rare attack types, further reducing its predictive performance in practical applications, especially in scenarios requiring accurate identification and response to rare threats. The ensemble approach involves multiple classifiers and optimization steps (such as PSO and LUS), which increases computational complexity and training time. This is impractical for real-time intrusion detection, where rapid response is crucial. The DeepFed model relies on federated learning, which aggregates model parameters from multiple industrial CPSs to build a collective intrusion detection model. While this approach preserves data privacy, it can lead to inconsistencies and variability in model performance due to the heterogeneous nature of data sources, as data quality and distribution can vary significantly among different industrial agents. DeepFed employs a CNN combined with a GRU. Although CNNs are effective in feature extraction and GRUs in handling sequential data, the model's architecture might not fully capture complex, high-dimensional, or less structured data patterns, especially when the data involves intricate temporal relationships or multidimensional anomalies that are common in cyber-physical system threats.

As shown in the tables, our model outperforms other models in performance metrics. This advantage is attributed to the fact that other models did not prioritize key features when dealing with high-dimensional features, resulting in less impressive performance in multi-class tasks. Additionally, the dataset exhibits class imbalance, with too few samples of rare classes, hindering the model’s ability to effectively learn critical features of these rare samples. Although SEL employs incremental learning to update the model by integrating historical knowledge, this approach may lead to inadequate adaptability when facing new types of attacks. Moreover, the efficiency of sustainable incremental learning models relies on updating individual classifiers and reusing historical knowledge, which can be computationally expensive. This makes them unsuitable for higher-efficiency operations in high-throughput environments like 5G and 6G networks. In summary, the SEL ensemble model does not adequately consider classifier sensitivity to different types of attacks and the issue of classifier weighting, resulting in lower accuracy.

Our approach considers classifier sensitivity to various attack types and assigns different weights to different attack types. Additionally, we incorporate dense connections and self-attention mechanisms, enabling the model to focus on and reuse important features. This is crucial in deep learning models, as without sufficient reuse mechanisms, key features may diminish as the network depth increases. Feature reuse prevents the loss of critical information by maintaining early-layer feature information. Consequently, we conclude that our model fully considers the sensitivity of individual classifiers to various types of attacks and the reuse of important features, demonstrating greater flexibility and adaptability when generating decision results. Thus, our model’s predictive capabilities surpass those of other models.

\begin{table*}[h!]
\centering
\renewcommand{\arraystretch}{1}
\caption{Performance Comparison of the Proposed Model and Existing State-of-the-Art Models on the CIC-UEBA Dataset for Intrusion Detection Task}
\fontsize{10}{12}\selectfont
\vspace{0.2cm}
\begin{tabular}{>{\centering\arraybackslash}p{2cm} >{\centering\arraybackslash}p{2cm} >{\centering\arraybackslash}p{2cm} >{\centering\arraybackslash}p{2cm}>{\centering\arraybackslash}p{2cm}>{\centering\arraybackslash}p{2cm}}
\toprule[0.8pt] 
\textbf{Classification} & \textbf{Metric} & \textbf{SEL} & \textbf{SKP} & \textbf{Ours} & \textbf{DeepFed} \\
\midrule[0.8pt] 
\multirow{4}{*}{DoS}   & Accuracy & 0.8923 & 0.8204  & \textbf{0.9763} & 0.8756 \\
                           & Precision & 0.8963 & 0.9107 & \textbf{0.9773} & 0.9411 \\
                                 & Recall    & 0.8364 & 0.9487 & \textbf{0.9865} & 0.8412 \\
                                 & F1-Score  & 0.8328 & 0.8865 & \textbf{0.9819} & 0.8679 \\
\midrule[0.8pt] 
\multirow{4}{*}{Probe}           & Accuracy & 0.9210 & 0.9231  & \textbf{0.9569} & 0.9402 \\
                                 & Precision & 0.9138 & 0.8725 & \textbf{0.9992} & 0.8548 \\
                                 & Recall   & 0.9213 & 0.9980  & 0.9908 & 0.8653\\
                                 & F1-Score & 0.9557 & 0.9718  & \textbf{0.9950} & 0.9552 \\
\midrule[0.8pt] 
\multirow{4}{*}{Benign}          & Accuracy    & 0.9314 & 0.9214 & \textbf{0.9992} & 0.9593 \\
                                 & Precision   & 0.9652 & 0.9004 & \textbf{0.9975} & 0.9492 \\
                                 & Recall      & 0.8610 & 0.9253& \textbf{0.9980} & 0.9430\\
                                 & F1-Score    & 0.9361 & 0.8841 & \textbf{0.9978} & 0.8927 \\
\midrule[0.8pt] 
\multirow{4}{*}{U2R}             & Accuracy    & 0.8999 & 0.9482 & 0.9383 & 0.9429 \\
                                 & Precision   & 0.9256 & 0.9455 & \textbf{0.9939} & 0.9140 \\
                                 & Recall      & 0.9222 & 0.9381& \textbf{0.9513} & 0.8726\\
                                 & F1-Score    & 0.9514 & 0.8432 & \textbf{0.9721} & 0.8512 \\
\midrule[0.8pt] 
\multirow{4}{*}{R2L}             & Accuracy    & 0.8790 & 0.9011 & \textbf{0.9418} & 0.9447 \\
                                 & Precision   & 0.9331 & 0.8501 & \textbf{0.9795} & 0.9633 \\
                                 & Recall      & 0.8358 & 0.8710 & \textbf{0.9820} & 0.9314\\
                                 & F1-Score    & 0.8729 & 0.9509 & \textbf{0.9807} & 0.9235 \\
\bottomrule[0.8pt] 
\end{tabular}
\label{table5}
\end{table*}
\begin{table*}[h!]
\centering
\renewcommand{\arraystretch}{1}
\caption{Performance Comparison of the Proposed Model and Existing State-of-the-Art Models on the KDD-UEBA Dataset for Intrusion Detection Task}
\fontsize{10}{12}\selectfont
\vspace{0.2cm}
\begin{tabular}{>{\centering\arraybackslash}p{2cm} >{\centering\arraybackslash}p{2cm} >{\centering\arraybackslash}p{2cm} >{\centering\arraybackslash}p{2cm}>{\centering\arraybackslash}p{2cm}>{\centering\arraybackslash}p{2cm}}
\toprule[0.8pt]
\textbf{Classification} & \textbf{Metric} & \textbf{SEL} & \textbf{SKP} & \textbf{Ours} & \textbf{DeepFed} \\
\midrule[0.8pt]
\multirow{4}{*}{DoS}             & Accuracy & 0.8635 & 0.8750 & \textbf{0.9948} & 0.7753 \\
                                 & Precision & 0.5860 & 0.8824 & \textbf{0.9773} & 0.9051\\
                                 & Recall   & 0.8892  & 0.9811  & \textbf{0.9865} & 0.9786 \\
                                 & F1-Score & 0.9074 & 0.7701 & \textbf{0.9819} & 0.9380 \\
\midrule[0.8pt]
\multirow{4}{*}{Probe}           & Accuracy & 0.8925 & 0.8792 & \textbf{0.9952} & 0.9201 \\
                                 & Precision & 0.9138 & 0.8643 & \textbf{0.9992} & 0.8430 \\
                                 & Recall   & 0.9782 & 0.9225 & \textbf{0.9908} & 0.8553\\
                                 & F1-Score & 0.9341 & 0.7086 & \textbf{0.9950} & 0.9018 \\
\midrule[0.8pt]
\multirow{4}{*}{Benign}          & Accuracy    & 0.7130 & 0.7086 & \textbf{0.9724} & 0.9933 \\
                                 & Precision   & 0.8342 & 0.7738 & \textbf{0.9975} & 0.9872 \\
                                 & Recall      & 0.8740 & 0.8741 & \textbf{0.9980} & 0.9738\\
                                 & F1-Score    & 0.9905 & 0.8301 & \textbf{0.9978} & 0.9415 \\
\midrule[0.8pt]
\multirow{4}{*}{U2R}             & Accuracy    & 0.9904 & 0.8823 & 0.9670 & 0.7741 \\
                                 & Precision   & 0.9043 & 0.9044 & \textbf{0.9939} & 0.7380 \\
                                 & Recall      & 0.8952 & 0.9001 & 0.9513 & 0.9203\\
                                 & F1-Score    & 0.8736 & 0.8367 & \textbf{0.9721 }& 0.8033 \\
\midrule[0.8pt]
\multirow{4}{*}{R2L}             & Accuracy    & 0.8324 & 0.7753 & \textbf{0.9620} & 0.9600 \\
                                 & Precision   & 0.8561 & 0.8349 & \textbf{0.9795} & 0.9538 \\
                                 & Recall      & 0.8299 & 0.8825 & \textbf{0.9820} & 0.9227\\
                                 & F1-Score    & 0.8843 & 0.9746 & \textbf{0.9807} & 0.8000 \\
\bottomrule[0.8pt]
\end{tabular}
\label{KDD-UEBA1}
\end{table*}
\subsubsection{Analysis of Insider Detection Effectiveness}
We selected the existing classification models showing superior performance in detecting insider threats for comparison. Our method performs well in four evaluation metrics, indicating that it achieves the best performance in addressing insider threats, as can be seen from Table \ref{TT1}.

Temporal features of user behavior are deemed significant in the analysis of user activities due to their ability to reflect the temporal characteristic of behavior patterns, provide predictive information, reveal anomalous actions, and enhance context awareness. However, the SVM model fails to consider these temporal characteristics, resulting in lower detection efficiency. SVM and LR are both linear models, which inherently limits their ability to handle complex nonlinear data effectively. Although SVM can address certain nonlinear problems through the use of kernel functions, its performance is highly dependent on the appropriate selection of the kernel and precise parameter tuning. Additionally, SVM's high computational complexity poses challenges, particularly in large-scale datasets. LR, on the other hand, operates under the assumption that features are independent and have significant linear relationships. In real-world applications, however, features are often interrelated and exhibit complex nonlinear relationships, which restricts LR's ability to capture deeper data patterns, consequently affecting its classification accuracy.

Furthermore, SVM is notably sensitive to noise, and when applied to large datasets containing noise, it has a tendency to overfit to these noisy data points, thus reducing its generalization ability. LR also shares a similar sensitivity to noise, and its simplistic assumptions make it inadequate for learning complex data patterns. These inherent limitations render SVM and LR less effective in managing complex, nonlinear, and high-dimensional datasets when compared to more advanced models, such as deep learning algorithms. The subpar performance of SVM and LR is therefore largely due to their limited capacity to capture nonlinear features and adapt to high-dimensional data environments.

The DualNet-CV address this issue by emphasizing the importance of time-series features. Nevertheless, DualNet-CV does not adequately consider the accumulation and reuse of knowledge from historical models. Cyber attacks are dynamic and evolve over time; therefore, detection models must be continuously updated to counter new threats. DualNet-CV underperforms compared to DenseAttDNN in insider threat detection, primarily due to its limitations in feature extraction, data augmentation, and behavior context processing. DualNet-CV relies on traditional feature extraction methods, encoding user behavior into image representations but lacks in-depth modeling of relationships between features, making it less effective in capturing complex user behavior patterns. In contrast, DenseAttDNN leverages attention mechanisms that focus on critical features within the input data, distinguishing normal from malicious behavior more effectively. The data augmentation strategy of DualNet-CV is also basic, utilizing simple image flips and rotations, which inadequately addresses class imbalance in the dataset. DenseAttDNN, on the other hand, employs advanced techniques such as adaptive synthetic sampling, significantly enhancing its detection capability for minority class malicious behaviors. Moreover, DualNet-CV struggles with temporal and contextual dependencies of user behavior, failing to effectively handle complex time-dependent threats. Its dual-input architecture merely concatenates image encodings with non-dynamic information, lacking a deep integration mechanism to utilize non-dynamic attributes (e.g., roles, psychological traits) to aid behavior judgment. These architectural limitations result in DualNet-CV's inferior performance in handling complex user entity behavior analysis tasks.

In the detection of insider threats, it is crucial to focus on temporal characteristics and user behavior features. User behavior features include, but are not limited to, users' login habits, email usage behaviors, and data download and upload activities. However, the plethora of features can prevent the model from correctly identifying genuinely useful information, leading to decreased predictive performance. Unlike the aforementioned methods, we employ the dot-product attention mechanism to enable the model to concentrate more on significant features while considering the accumulation of historical behaviors. This mechanism helps the model focus on features most relevant to security threats, such as abnormal access patterns or unusual data transmission behaviors, enabling more accurate identification of key user or entity behavioral patterns. Consequently, on the datasets we proposed, our solution demonstrates a clear advantage over the other three models across all metrics.

\begin{table*}[h!]
\centering
\caption{Performance Comparison of the Proposed Model and Existing Models for Insider Threat Detection Task}
\fontsize{10}{12}\selectfont
\renewcommand{\arraystretch}{1}
\begin{tabular}{>{\centering\arraybackslash}p{3cm} >{\centering\arraybackslash}p{2cm} >{\centering\arraybackslash}p{2cm} >{\centering\arraybackslash}p{2cm}>{\centering\arraybackslash}p{2cm}>{\centering\arraybackslash}p{2cm}>{\centering\arraybackslash}p{2cm}}
\toprule[0.8pt]
\textbf{Dataset}       & \textbf{Classification} & \textbf{Metric}  & \textbf{SVM} & \textbf{LR} & \textbf{Ours} & \textbf{DualNet-CV}\\
\midrule[0.8pt]
 \multirow{8}{*}{KDD-UEBA} & \multirow{4}{*}{Normal}  & Accuracy\setlength{\baselineskip}{15pt}   & 0.7800 & 0.8185 & \textbf{0.9890} & 0.8324\\
                           & \multirow{4}{*}{User}   & Precision\setlength{\baselineskip}{15pt}   & 0.8024 & 0.8143 & \textbf{0.9642} & 0.9022\\
                           &                          & Recall\setlength{\baselineskip}{15pt}     & 0.8110 & 0.9104 & \textbf{0.9890} & 0.9212\\
                           &                          & F1-Score\setlength{\baselineskip}{15pt}   & 0.8002 & 0.8221 & \textbf{0.9764} & 0.8203\\
\cmidrule(lr){2-7}
                           & \multirow{4}{*}{Malicious} & Accuracy  & 0.8002 & 0.8031 &\textbf{0.9617} & 0.8153\\
                           & \multirow{4}{*}{User}     & Precision  & 0.8099 & 0.7823 &\textbf{0.9919} & 0.8422\\
                           &                          & Recall      & 0.8037 & 0.8120 &\textbf{0.9044} & 0.8931\\
                           &                          & F1-Score    & 0.8122 & 0.7913 & \textbf{0.9462} & 0.8435\\
\midrule[0.8pt]
\multirow{8}{*}{KDD-UNBLogs} & \multirow{4}{*}{Normal} & Accuracy    & 0.8151 & 0.8207 & \textbf{0.9044}  & 0.8410\\
                           &  \multirow{4}{*}{User}    & Precision   & 0.7841 & 0.8103 & \textbf{0.9364} & 0.9002\\
                           &                          & Recall      & 0.7812 & 0.7782 & \textbf{0.9557} & 0.8510\\
                           &                          & F1-Score    & 0.8007 & 0.8223 & \textbf{0.9460} & 0.7800\\
\cmidrule(lr){2-7}
                           & \multirow{4}{*}{Malicious} & Accuracy    & 0.8130 & 0.8145 & \textbf{0.9715} & 0.8123\\
                           & \multirow{4}{*}{User}    & Precision   & 0.8124 & 0.8000 & \textbf{0.9892} & 0.8324\\
                           &                          & Recall     & 0.8125 & 0.8190 &\textbf{0.9891} & 0.8022\\
                           &                          & F1-Score  & 0.8231 & 0.8415 & \textbf{0.9891} & 0.8002\\
\bottomrule[0.8pt]
\end{tabular}%
\label{TT1}
\end{table*}

\begin{table*}[h!]
\centering
\caption{Performance Comparison of the Proposed Model and Existing Models for Synergistic Detection Task}
\fontsize{10}{12}\selectfont
\renewcommand{\arraystretch}{1}
\begin{tabular}{>{\centering\arraybackslash}p{3cm} >{\centering\arraybackslash}p{2cm} >{\centering\arraybackslash}p{2cm} >{\centering\arraybackslash}p{2cm}>{\centering\arraybackslash}p{2cm}>{\centering\arraybackslash}p{2cm}>{\centering\arraybackslash}p{2cm}}
\toprule[0.8pt] 
\textbf{Dataset}       & \textbf{Classification} & \textbf{Metric}  & \textbf{SVM} & \textbf{LR} & \textbf{Ours} & \textbf{DualNet-CV}\\
\midrule[0.8pt] 
                            & \multirow{4}{*}{Normal}  & Accuracy\setlength{\baselineskip}{15pt}      & 0.7423 & 0.8065 & \textbf{1.0000} & 0.9120\\
                           & \multirow{4}{*}{User}     & Precision\setlength{\baselineskip}{15pt}   & 0.7862 & 0.8437 & \textbf{0.9992}  & 0.9435\\
                           &                          & Recall\setlength{\baselineskip}{15pt}      & 0.8096 & 0.8771 & \textbf{1.0000 }& 0.9153\\
\multirow{8}[14]{*}{KDD-UEBA}  &                          & F1-Score\setlength{\baselineskip}{15pt}    & 0.7741 & 0.8556 & \textbf{0.9996} & 0.9311\\
\cmidrule(lr){2-7}
                           & \multirow{4}{*}{Malicious}    & Accuracy  & 0.7380 & 0.8096 & \textbf{0.9676}  & 0.9318\\
                           & \multirow{4}{*}{User}     & Precision   & 0.7982 & 0.8743 & \textbf{1.0000} & 0.9280\\
                           &                           & Recall      & 0.8033 & 0.8830 & \textbf{0.9676} &0.8877\\
                           &                          & F1-Score    & 0.8411 & 0.9000 & \textbf{0.9836} &0.9003\\
\cmidrule(lr){2-7}
                           & \multirow{4}{*}{Intruder}   & Accuracy    & -- & -- & \textbf{0.9788} & --\\
                           &  \multirow{4}{*}{    }    & Precision   & -- & -- & \textbf{1.0000} & --\\
                           &                          & Recall      & -- & -- & \textbf{0.9788} & --\\
                           &                          & F1-Score    & -- & -- & \textbf{0.9893} & --\\
\cmidrule(lr){2-7}
                           & \multirow{4}{*}{Potential}  & Accuracy    & -- & -- & \textbf{0.9960} & --\\
                           & \multirow{4}{*}{Intruder}      & Precision   & -- & -- & \textbf{0.9973 }& --\\
                           &                          & Recall      & -- & -- &\textbf{ 0.9960} & --\\
                           &                          & F1-Score    & -- & -- & \textbf{0.9967} & --\\

\midrule[0.8pt] 
                           & \multirow{4}{*}{Normal}   & Accuracy    & 0.7910 & 0.7770 &\textbf{0.9999} & 0.8762\\
                           & \multirow{4}{*}{User}          & Precision   & 0.8132& 0.8906 & \textbf{1.0000} &0.8650\\
                           &                          & Recall      & 0.7775 & 0.7736& \textbf{1.0000} &0.8910\\
\multirow{8}[14]{*}{NSL-UEBA}       &                          & F1-Score    & 0.7863& 0.7417 &\textbf{1.0000}& 0.8415\\
\cmidrule(lr){2-7}
                           & \multirow{4}{*}{Malicious}  & Accuracy    & 0.7354 & 0.7436 & \textbf{0.9382}&0.7810\\
                           & \multirow{4}{*}{User}      & Precision   & 0.8096 & 0.7849 & \textbf{0.9815} &0.1253\\
                           &                           & Recall & 0.8254 & 0.8559 & \textbf{0.9382} & 0.7563\\
                           &                          & F1-Score    & 0.8003 & 0.8251& \textbf{0.9594} &0.7753\\
\cmidrule(lr){2-7}
                           & \multirow{4}{*}{Intruder}  & Accuracy    & -- & -- & \textbf{0.9877}& --\\
                           & \multirow{4}{*}{   }      & Precision   & -- & -- &\textbf{ 0.9878} & --\\
                           &                          & Recall      & -- & -- & \textbf{0.9878} & --\\
                           &                          & F1-Score    & -- & -- & \textbf{0.9878} & --\\
\cmidrule(lr){2-7}
                           & \multirow{4}{*}{Potential}  & Accuracy    & -- & -- & \textbf{0.9960} & --\\
                           & \multirow{4}{*}{Intruder}      & Precision   & -- & -- &\textbf{0.9972} & --\\
                           &                           & Recall      & -- & -- & \textbf{0.9961} & --\\
                           &                          & F1-Score    & -- & -- & \textbf{0.9966} & --\\
                           
\bottomrule[0.8pt] 
\end{tabular}%
\label{tab9}
\end{table*}

\subsubsection{Analysis of Synergistic Detection Effectiveness}
In our study, we conducted a comprehensive evaluation of the model's capability to detect attacks using two synthetic datasets, as detailed in Table \ref{tab9}. Notably, the '-' symbol in the table indicates that the model was unable to detect the threats. The experimental outcomes affirm that the model exhibits high efficacy in detecting a broad spectrum of attacks, particularly rare attack types. The proposed Synergistic Detection Framework plays a crucial role not only in unifying the detection of both insider and external threats within a single model, thereby enhancing resource utilization efficiency and detection efficiency, but also in effectively identifying intruder and potential intruder that traditional UEBA struggle to detect.

While traditional UEBA technologies often struggle with identifying masquerader attacks due to their reliance on user activities or device usage analysis, the DenseAttDNN model offers a more nuanced approach.  It provides a sophisticated means of analyzing behavioral patterns with Several perspectives: Temporal Behavioral Pattern, Sequential Behavioral Pattern, and Contextual Behavioral Pattern.

To effectively analyze and detect unusual behavioral patterns, our approach centers on dense connectivity and attention mechanisms, which enable the network to capture features from various hierarchical levels, thereby encompassing a broader range of behavioral characteristics. Each layer receives the concatenated outputs of all preceding layers, which are processed through nonlinear activation functions to generate features for further behavioral analysis. This dense connectivity helps DenseAttDNN effectively mitigate the vanishing gradient problem, allowing learning signals to propagate smoothly through the network and thus supporting the training of deep architectures. The self-attention mechanism in DenseAttDNN adds further flexibility and precision to behavioral analysis. For each layer, this mechanism allows the network to dynamically adjust the importance of each feature based on the entire feature set’s computed weights. This capability enables DenseAttDNN not only to analyze periodicity and regularity in time-series data but also to conduct complex assessments of user behavior over time. By leveraging temporal features, the model can identify temporal anomalies when users deviate from their typical schedules, such as accessing the system during unusual hours or performing specific actions at abnormal frequencies. These irregularities may indicate compromised accounts or insider threats, where attackers exploit the system outside regular hours to evade detection. Furthermore, DenseAttDNN considers the sequence of actions performed, which is critical for capturing dependencies in behavior. By analyzing the order of actions, the model identifies normal sequences of operations when users access specific systems. Deviations from these established patterns, such as directly accessing sensitive data while bypassing intermediate steps, can signal potential security breaches. This keen awareness of sequential patterns enables DenseAttDNN to identify possible violations early.

Contextual factors further enhance detection capabilities by evaluating the environment in which actions occur, including device types, locations, and network connections. The weights calculated by the self-attention mechanism can dynamically adjust based on these contextual features, allowing the model to assess user behavior within the appropriate environmental context. For instance, if a user typically logs in from a consistent location but suddenly logs in from a distant, unusual location or uses a new device not previously associated with them, these contextual shifts may suggest compromised credentials or inappropriate device usage.

Additionally, our approach incorporates resampling techniques to increase the representation of anomalous behaviors in the dataset, which enhances DenseAttDNN's ability to learn from these outliers and improves its overall classification performance. For example, if an employee in a specific role suddenly begins accessing systems outside their typical behavior cluster, this could indicate an insider threat or a compromised account. By comparing individual behaviors against peer norms—such as typical login times and locations among employees in the same department—the model can detect significant deviations that may suggest malicious activities or unauthorized access.

This comprehensive analysis of behavioral patterns, integrating insights from temporal, sequential, and contextual dimensions, provides a robust framework for detecting masquerader attacks. By seamlessly integrating various aspects of user behavior, our approach significantly enhances system security against unauthorized and anomalous activities. By analysing the above aspects, DenseAttDNN enhances the detection of both insider and external threats.

\begin{figure*}[ht]
\centering
\begin{subfigure}[b]{0.24\linewidth}
    \includegraphics[width=\linewidth]{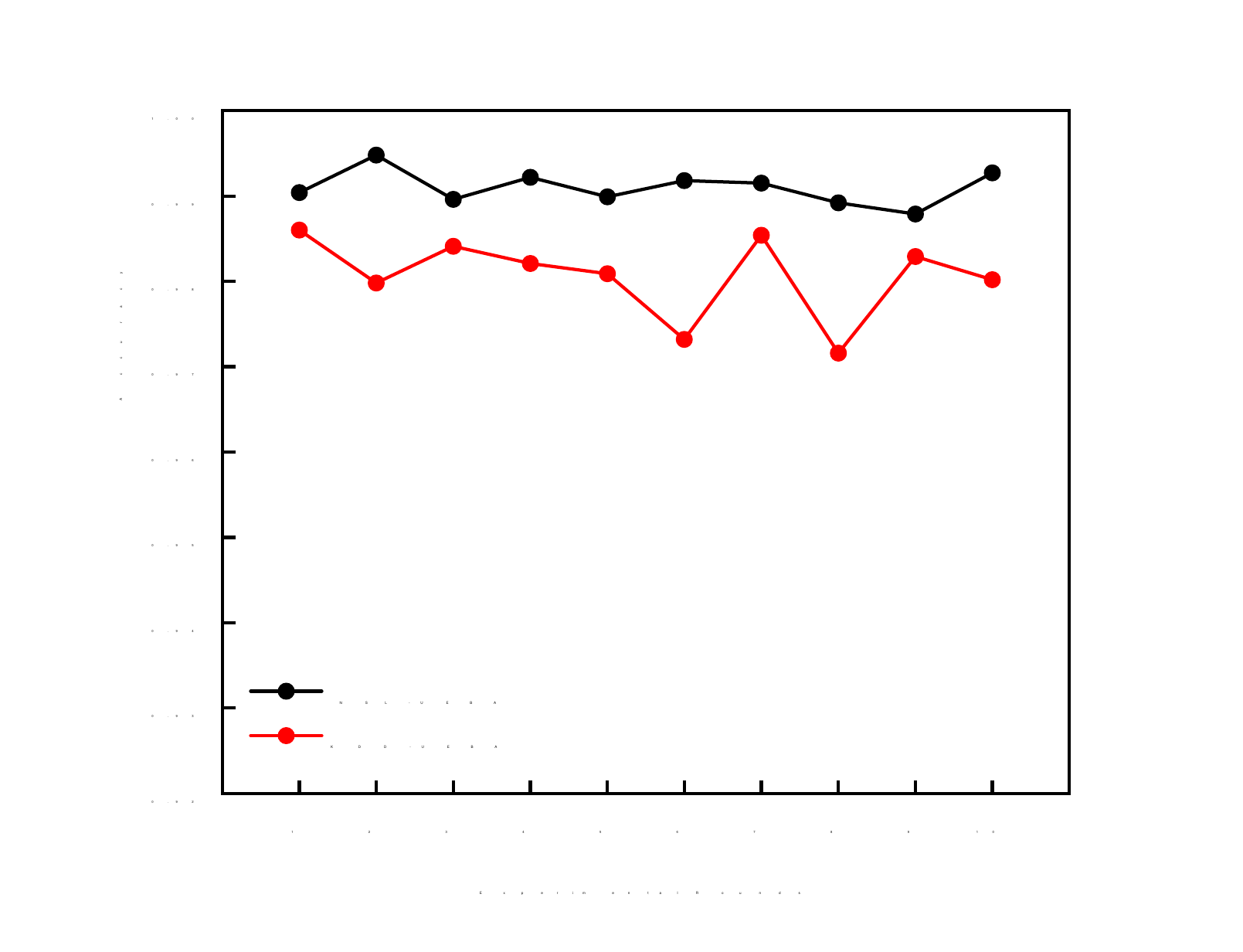}
    \caption{Acc Test Statistics}
    \label{fig:acc-test}
\end{subfigure}%
\begin{subfigure}[b]{0.24\linewidth}
    \includegraphics[width=\linewidth]{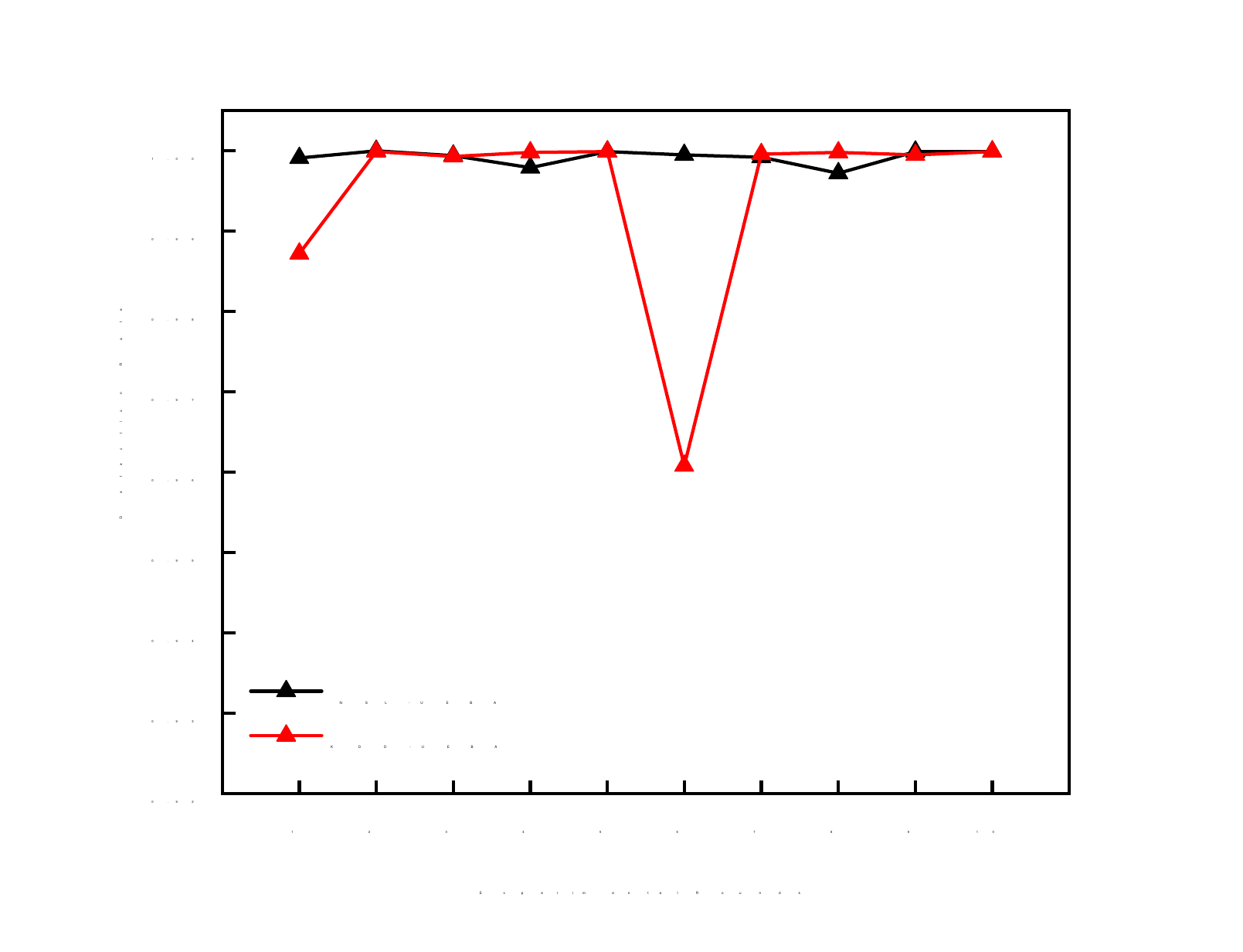}
    \caption{DR Test Statistics}
    \label{fig:dr-test}
\end{subfigure}%
\begin{subfigure}[b]{0.24\linewidth}
    \includegraphics[width=\linewidth]{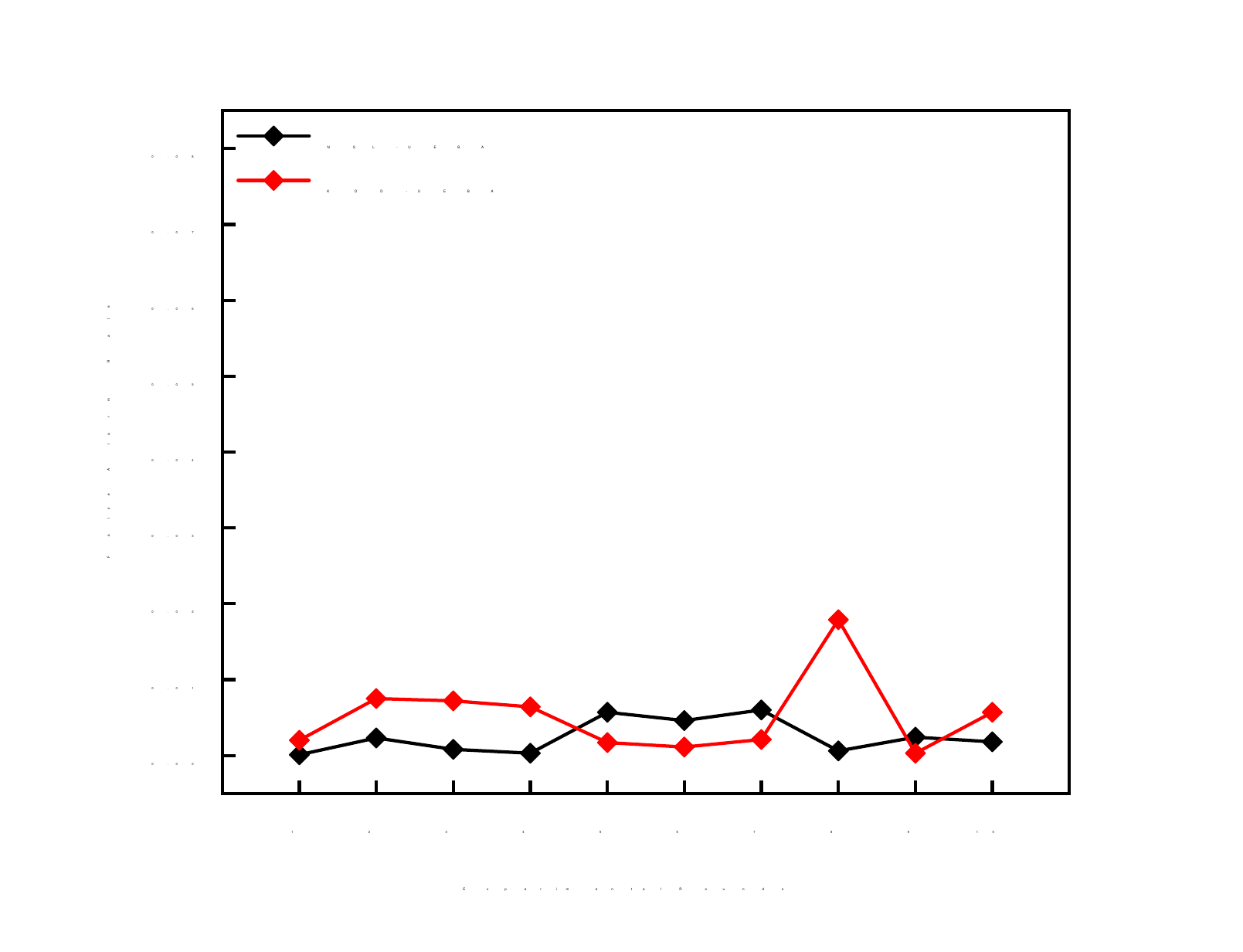}
    \caption{FAR Test Statistics}
    \label{fig:far-test}
\end{subfigure}%
\begin{subfigure}[b]{0.24\linewidth}
    \includegraphics[width=\linewidth]{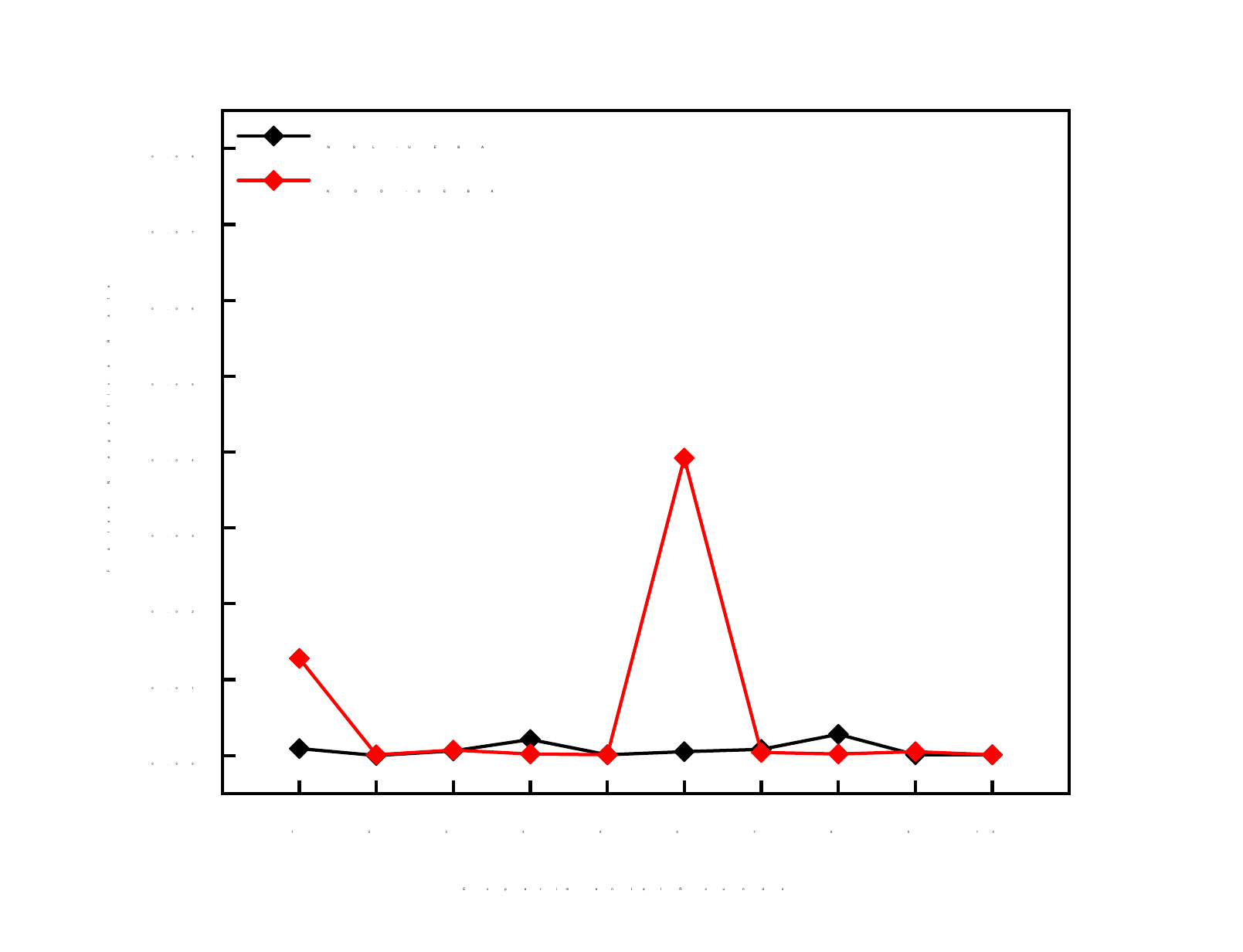}
    \caption{FNR Test Statistics}
    \label{fig:fnr-test}
\end{subfigure}

\begin{subfigure}[b]{0.24\linewidth}
    \includegraphics[width=\linewidth]{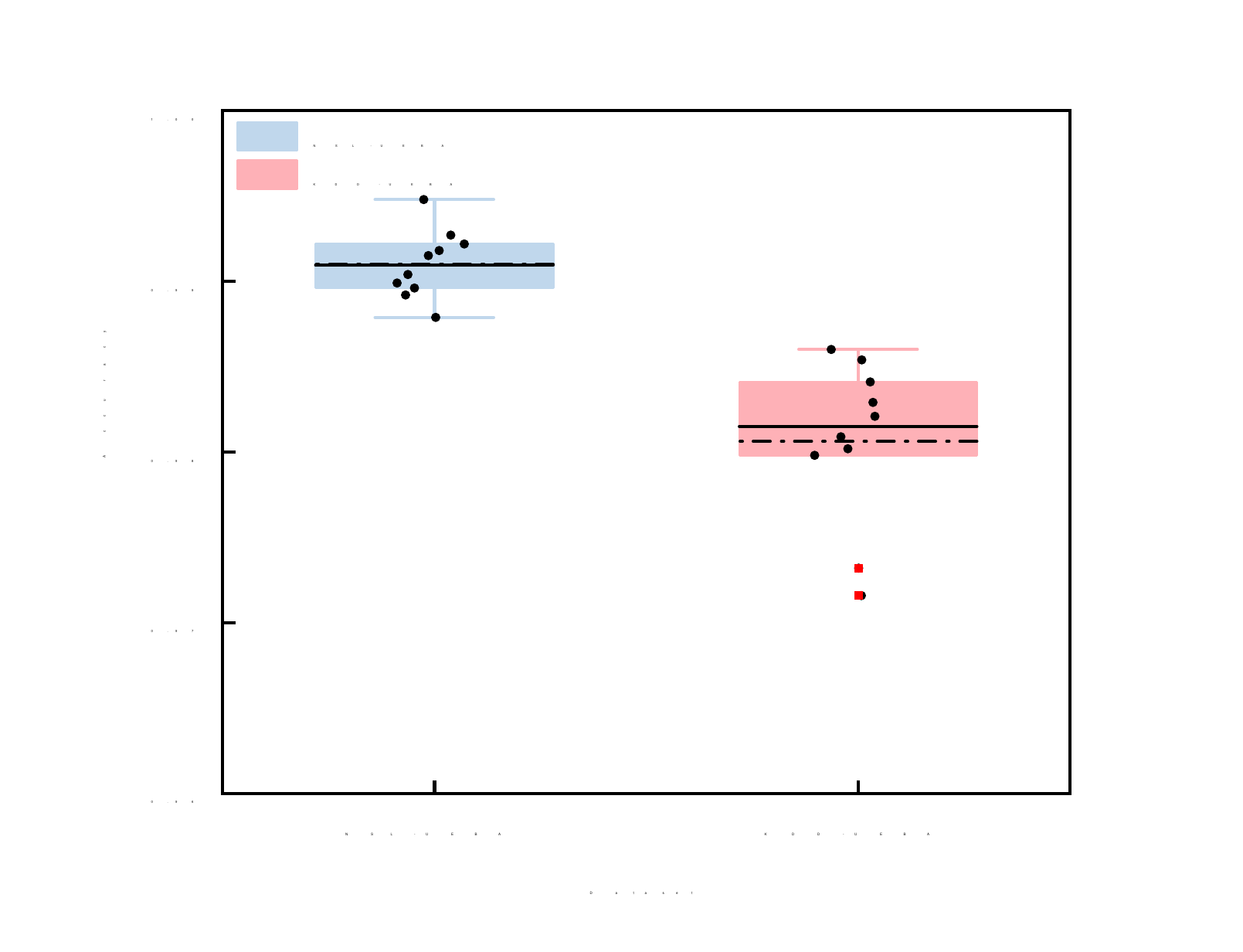}
    \caption{Acc Test Analysis}
    \label{fig:acc-analysis}
\end{subfigure}%
\begin{subfigure}[b]{0.24\linewidth}
    \includegraphics[width=\linewidth]{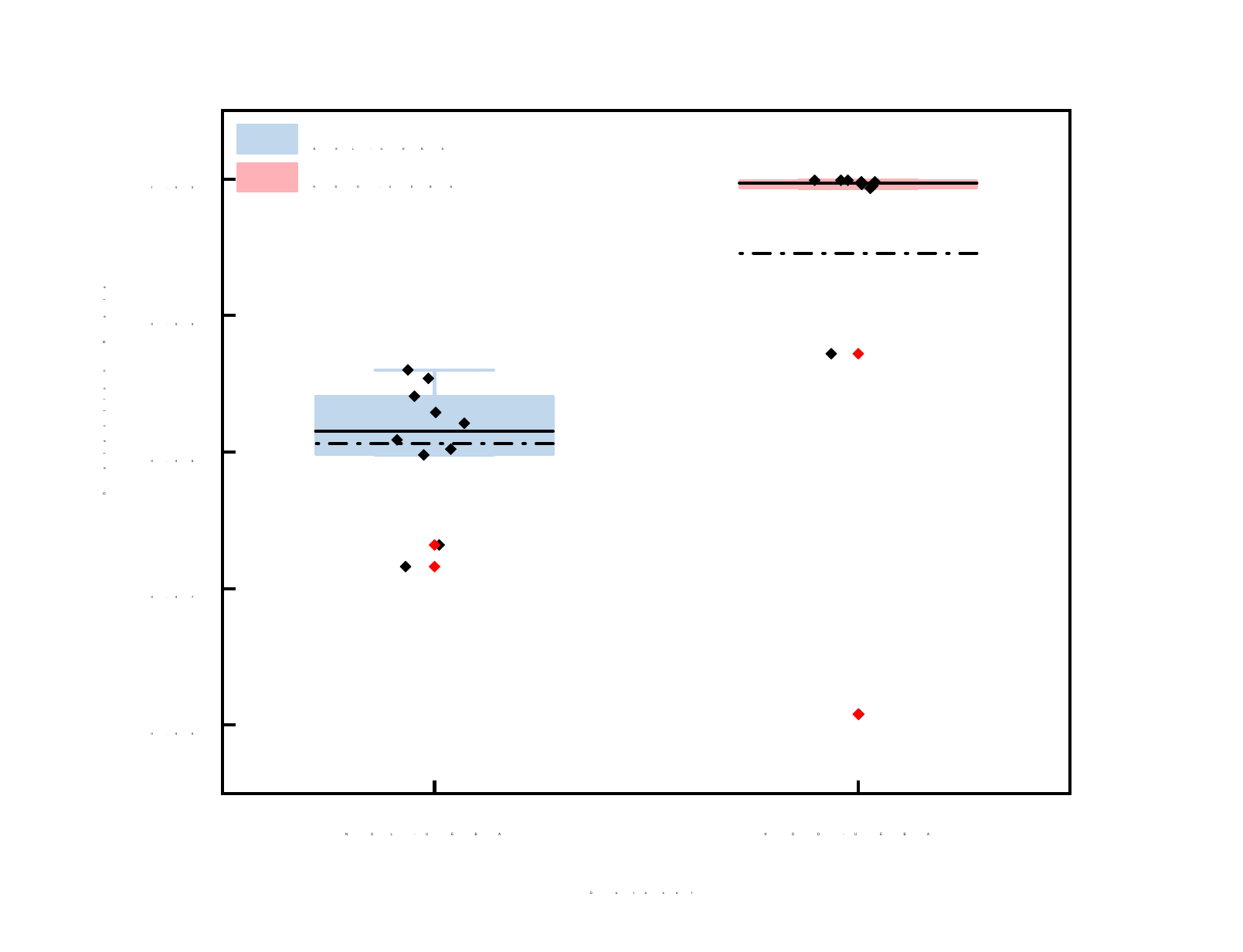}
    \caption{DR Test Analysis}
    \label{fig:dr-analysis}
\end{subfigure}%
\begin{subfigure}[b]{0.24\linewidth}
    \includegraphics[width=\linewidth]{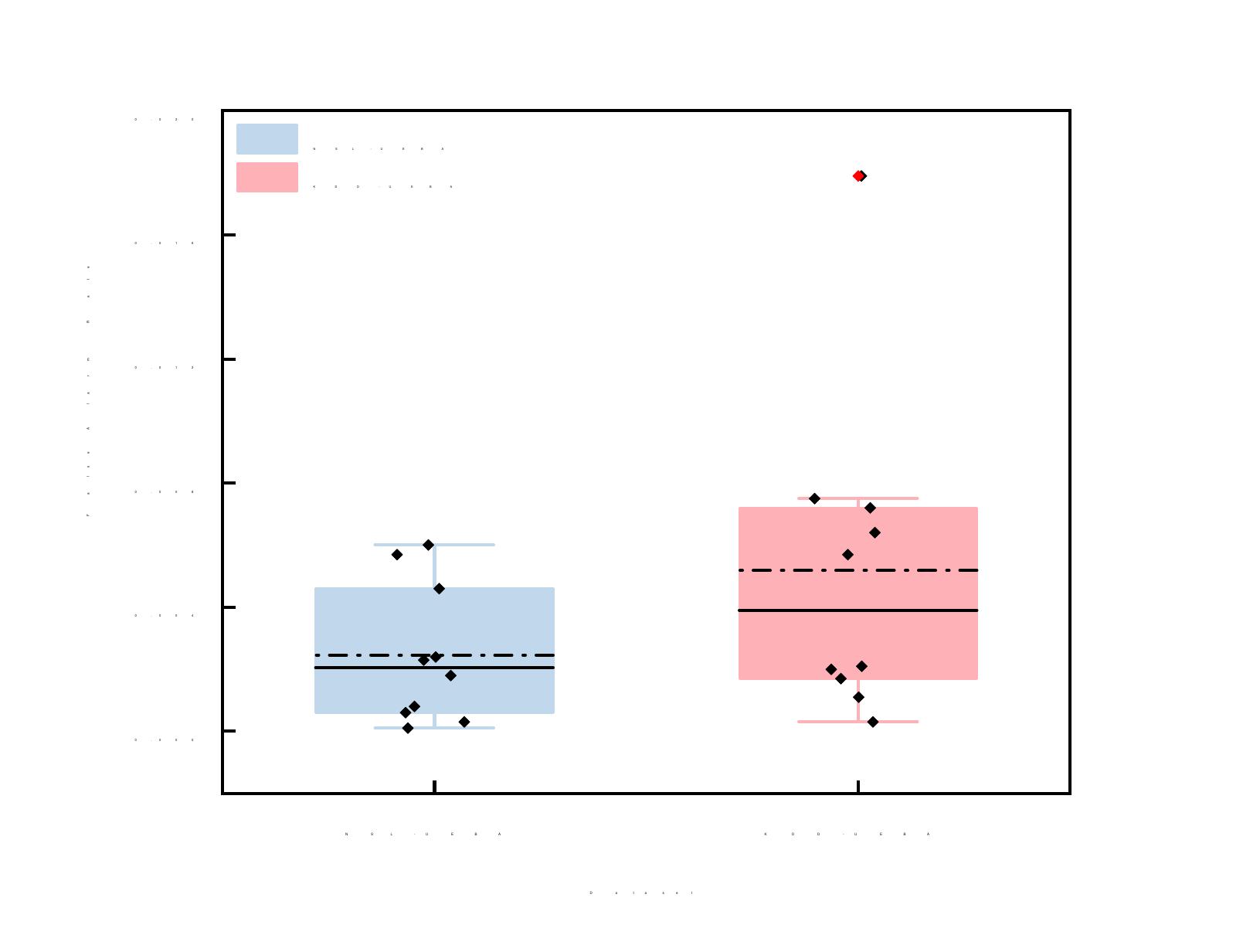}
    \caption{FAR Test Analysis}
    \label{fig:far-analysis}
\end{subfigure}%
\begin{subfigure}[b]{0.24\linewidth}
    \includegraphics[width=\linewidth]{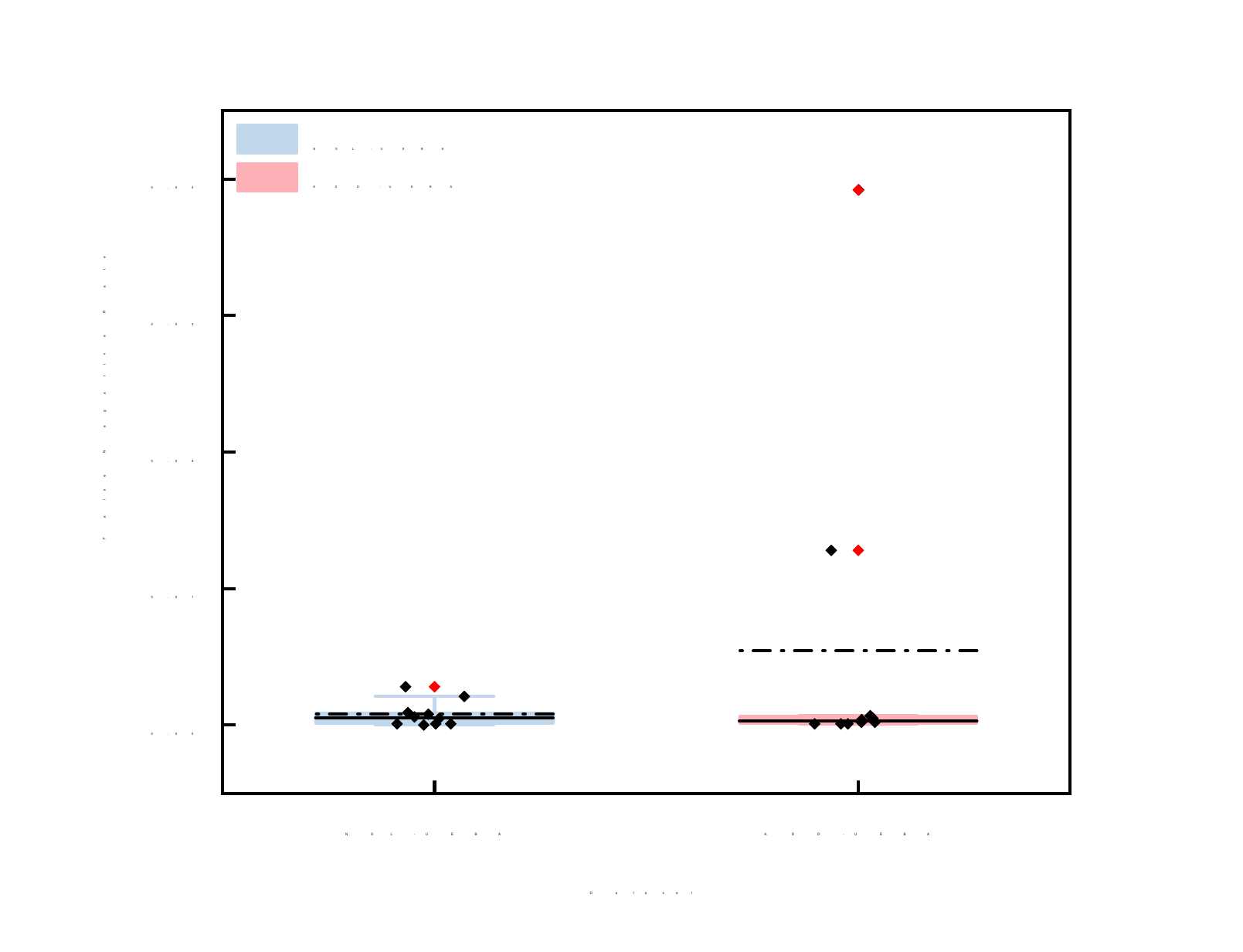}
    \caption{FNR Test Analysis}
    \label{fig:fnr-analysis}
\end{subfigure}

\caption{Test Statistics and Analysis}
\label{fig7}
\end{figure*}

\subsection{Stability Analysis}

To validate the robustness and stability of our model in practical threat detection scenarios, we conducted ten repeated experiments on the NSL-UEBA and KDD-UEBA datasets.  We recorded key performance metrics, including Accuracy, DR, FAR, and FNR.  Fig. \ref{fig7} illustrates the trends of these metrics across the repeated experiments, demonstrating consistent performance with minimal variability.

The line plots (Figures 5a-5d) depict the performance trends over the experimental rounds, showing that our model maintains stable and high accuracy while keeping FAR and FNR low.  To further emphasize the consistency of our results, we used box plots (Figures 5e-5h) to provide a statistical analysis of the outcomes.

In the box plots, the solid black line represents the median, the dashed black line indicates the mean, and the solid black dots show the distribution of the test data.  Outliers are highlighted with red solid dots, while the red solid lines, representing the whiskers, show the range of non-outlier data.  These visualizations confirm that our model exhibits steady performance with limited deviation, reinforcing its reliability in threat detection applications.

\begin{itemize}
    \item \textbf{Accuracy Analysis.} The accuracy rates for both datasets remained above 0.96, indicating that the model has a high baseline level of accuracy. Moreover, with the median values being above 0.98, it demonstrates that the model typically achieves a high degree of accuracy. The combination of high median accuracy and narrow interquartile ranges, particularly in the NSL-UEBA, signifies that the model maintains stable performance across different experiments.
\end{itemize}

\begin{itemize}
    \item \textbf{Detection Rate Analysis.} While there was significant fluctuation in the KDD-UEBA dataset during the sixth round of experiments, its rapid recovery showcases the model's capability to rebound from anomalies, which is a critical aspect of stability. Overall, the median and interquartile range of detection rates across both datasets indicate a high degree of stability in the model. Although there are several outliers in the KDD-UEBA dataset, they do not substantially impact the overall stability.
\end{itemize}
\begin{itemize}
    \item \textbf{False Alarm Rate Analysis.} In the majority of experimental runs, the false alarm rate for both datasets was maintained at a relatively low level. This indicates that, in most cases, the model is able to reliably differentiate between normal and anomalous activities, demonstrating good stability. Despite some fluctuations, the range of variation is not wide, and the overall low false alarm rate of the model signifies its high credibility and stability in detecting potential threats.
\end{itemize}
\begin{itemize}
    \item \textbf{False Negative Rate Analysis.} The model maintains a low false negative rate, meaning it correctly identifies threats in most instances. This demonstrates that the model can reliably detect threats, and despite fluctuations, it typically maintains stability. These results support the model's robustness and stability across different datasets and experimental conditions.
\end{itemize}

The analysis above indicate that our model exhibits high accuracy, high detection rates, low False Alarm Rates, and low False Negative Rates. This is attributable to the incorporation of attention mechanisms during the model's training phase, which enables the model to focus more on important features. Additionally, dense connections have been integrated into the hidden layers, allowing the model to efficiently pass significant feature information to the subsequent layer of neurons. It is precisely this focus on critical features that endows the model proposed in this paper with its characteristic high stability.
\linespread{1.5}
\begin{table*}[htbp]
  \centering
  \caption{Scalability Evaluation of Synergistic Detection using CIC-UEBA Dataset}
    \fontsize{10}{12}\selectfont
    \resizebox{\textwidth}{!}{
    \begin{tabular}{ccccccccccccc}
    \toprule[0.8pt] 
    \multirow{2}[4]{*}{\makecell{Dataset Size}} & \multicolumn{3}{c}{Latency(s)} & \multicolumn{3}{c}{Accuracy(\%)} & \multicolumn{3}{c}{Detection Time (s)} & \multicolumn{3}{c}{RAM Usage (\%)} \\
\cmidrule(lr){2-4} \cmidrule(lr){5-7} \cmidrule(lr){8-10} \cmidrule(lr){11-13}& XGB & \textbf{Ours} & GBDT & XGB & \textbf{Ours} & GBDT & XGB & \textbf{Ours} & GBDT & XGB & \textbf{Ours} & GBDT \\
    \midrule[0.8pt] 
    10\% & 0.2384   & {\color{green}{$\downarrow$}}\quad \textbf{0.1877}\quad {\color{blue}{$\approx$}}   & 0.1836   & 86.72\%   & {\color{red}{$\uparrow$}}\quad \textbf{98.73\%} \quad{\color{red}{$\uparrow$}}  & 88.81\%   & 450.867
 &{\color{green}{$\downarrow$}}\quad\textbf{294.522}\quad{\color{green}{$\downarrow$}}   & 384.792
 & 33.82\%
 &{\color{green}{$\downarrow$}}\quad\textbf{31.01}\%\quad{\color{green}{$\downarrow$}}   & 35.82\% \\
    25\% & 0.7234    & {\color{green}{$\downarrow$}}\quad \textbf{0.4731}\quad {\color{green}{$\downarrow$}}   & 0.7730   & 87.88\%   & {\color{red}{$\uparrow$}}\quad \textbf{99.52\%} \quad{\color{red}{$\uparrow$}}  & 89.21\%   & 789.326
 &{\color{green}{$\downarrow$}}\quad\textbf{742.247}\quad{\color{green}{$\downarrow$}}   & 798.239
 & 37.95\%
 &{\color{green}{$\downarrow$}}\quad\textbf{34.25}\%\quad{\color{green}{$\downarrow$}}   & 36.45\%
\\
    50\% & 0.9450    & {\color{green}{$\downarrow$}}\quad \textbf{0.9335}\quad {\color{green}{$\downarrow$}}   & 0.9653   & 90.32\%   &{\color{red}{$\uparrow$}}\quad \textbf{99.89\%} \quad {\color{red}{$\uparrow$}}   & 93.76\%   & 1034.38
 &{\color{red}{$\uparrow$}}\quad\textbf{1482.726
}\quad{\color{red}{$\uparrow$}}   & 730.241
 & 42.03\%
 &{\color{green}{$\downarrow$}}\quad\textbf{35.20}\%\quad{\color{red}{$\uparrow$}}   & 38.23\%\\
    75\% & 2.3196   & {\color{green}{$\downarrow$}}\quad \textbf{1.3922}\quad {\color{green}{$\downarrow$}}   & 1.5378   & 92.53\%   &{\color{red}{$\uparrow$}}\quad\textbf{99.69}\%\quad {\color{red}{$\uparrow$}}   & 95.52\%   & 2224.881   &{\color{blue}{$\approx$}}\quad\textbf{2226.399
}\quad {\color{green}{$\downarrow$}} & 2228.968
 & 42.88\%
 &{\color{green}{$\downarrow$}}\quad\textbf{37.96}\%\quad{\color{green}{$\downarrow$}}   & 35.77\% \\
    100\% & 4.5998   & {\color{green}{$\downarrow$}}\quad \textbf{1.7808}\quad {\color{green}{$\downarrow$}}   & 2.7963   & 94.55\%   &{\color{red}{$\uparrow$}}\quad\textbf{99.40}\%\quad{\color{red}{$\uparrow$}}& 96.00\%   & 3059.013
   & {\color{green}{$\downarrow$}}\quad\textbf{3023.048
}\quad {\color{green}{$\downarrow$}}& 4038.753
 & 43.53\%
 &{\color{green}{$\downarrow$}}\quad\textbf{39.88}\%\quad{\color{blue}{$\approx$}}   & 39.23\%
\\
    \bottomrule[0.8pt] 
    \end{tabular}%
    }
  \label{tt11}%
\end{table*}%
\subsection{Scalability Analysis}

Table \ref{tt11} presents a comparative evaluation of experimental results between the proposed framework, XGB, and GBDT using the CIC-UEBA dataset. In addition to accuracy and detection metrics, we also compared RAM usage, taking into account the relative frequency of computations. The bolded values in the table highlight the performance of our proposed framework, with the left arrow indicating a comparison with XGB and the right arrow indicating a comparison with GBDT. Red arrows signify an increase in the respective metric, green arrows indicate a decrease, and blue arrows denote stability, with no change in performance.

The results clearly demonstrate that DenseAttDNN significantly improves latency, detection accuracy, and detection time across all dataset sizes. Regarding RAM usage, DenseAttDNN also shows superior performance in most cases, reflecting its efficiency and scalability.
XGB and GBDT, as ensemble learning methods based on decision trees, demonstrate excellent performance in handling low-dimensional and medium-scale data by iteratively building trees to fit the residuals. However, as data dimensions and scale increase, the depth and number of trees grow rapidly, leading to significant computational and storage costs, especially when each tree's structure and split nodes require separate storage, resulting in exponential memory usage. Moreover, the process of generating decision trees exhibits high serial dependency, particularly during the optimization of split nodes, which requires layer-by-layer progression. Although XGB has introduced parallelization techniques such as column block splitting and approximate tree learning to alleviate this problem, it fundamentally remains constrained by the sequential nature of tree construction and its serialized dependencies, limiting its ability to fully leverage the parallelization capabilities of modern computing hardware. This architectural limitation results in a rapid increase in computational time and memory consumption as the dataset size grows. In contrast, DenseAttDNN fully capitalizes on modern parallel computing hardware, such as GPUs and TPUs, as its core operations (e.g., matrix multiplication and element-wise operations) are inherently suitable for parallel processing, allowing each layer to be computed simultaneously and significantly reducing training and inference time.

More importantly, DenseAttDNN achieves a qualitative leap in feature selection and representation capabilities through the integration of Self-Attention Mechanism and Dense Connections. The self-attention mechanism enables the model to dynamically allocate weights among different input features, allowing it to automatically focus on the most relevant feature subsets. This approach surpasses the static splitting method of XGB and GBDT's tree structures, effectively capturing long-range dependencies and complex feature interactions, which is particularly advantageous for handling large-scale and high-dimensional data. By performing weighted summation operations in the feature space, the attention mechanism reduces redundant computations, enhancing the model's representational power and training efficiency. Simultaneously, DenseAttDNN employs a dense connection architecture similar to DenseNet, where each layer receives inputs from all preceding layers, significantly enhancing feature reuse and gradient propagation efficiency while effectively mitigating the vanishing gradient problem. This architectural design ensures that features can be continuously refined and reused at deeper layers, substantially improving training efficiency and reducing dependency on data size. Therefore, the multi-layer feature fusion and dynamic feature selection capabilities of DenseAttDNN provide a robust foundation for its application in complex cybersecurity scenarios, showcasing superior scalability and adaptability.

\section{Limitation}
Currently, the main limitation of our framework is the lack of an automated mechanism for the dynamic recalibration of model weights. This deficiency impedes our ability to adaptively adjust weights across different categories of cyber threats, a critical process for upholding the defensive integrity of the system. This rigidity could potentially result in suboptimal threat detection and response capabilities. To address this challenge, significant enhancements to the framework are imperative. Future research will explore the development of advanced, self-evolving algorithms that autonomously optimize weight distributions in real-time, adapting to evolving cyber threats. Such advancements are expected to not only ameliorate existing limitations but also strengthen the system’s resilience against increasingly sophisticated and adaptive cyber threats. We will also considering something about the security of LLM services in wireless environment \cite{liu2024llm}.

\section{Conclusion}
As digital engagement continues to surge, individuals are increasingly willing to share personal information with organizations in exchange for personalized services and improved transactional efficiency. This evolving landscape highlights the paramount importance of implementing robust data protection measures, positioning them as a critical business priority amid the growing sophistication of cyber threats. To address the challenge of detecting increasingly complex attack vectors that blur the boundaries between insider and external threats. We propose an advanced, synergistic detection framework that seamlessly integrates IDS with UEBA technologies. In addition, to address the limitations in the breadth and depth of existing insider threat datasets, which hinder the detection capabilities of machine learning and deep learning models against diverse attack vectors, we introduce four meticulously curated, comprehensive datasets. These datasets encompass an extensive array of attack scenarios, specifically designed to bolster the detection prowess of advanced models. By providing a richer, more varied training foundation, these datasets aim to empower machine learning and deep learning models to effectively adapt to and mitigate a wider spectrum of emerging threats. The primary aim of this research is to enhance the detection of both insider and external threats, with a particular focus on identifying intruders and potential intruders that are often missed by traditional UEBA methods. By addressing these limitations, the proposed framework effectively strengthens the ability to detect increasingly sophisticated attack vectors, thereby safeguarding critical assets and enhancing overall security resilience. The proposed scheme integrates dense connections and self-attention mechanism, which enhances the model's ability to discern critical features crucial for detecting Internet-based attacks. Simulation experiments validate the effectiveness of the model, demonstrating its synergy and robustness, along with its capacity to identify rare attacks, thereby achieving the desired security outcomes. Performance evaluations reveal that our model achieves test accuracies of 98.96\% and 99.12\% on the KDD-UEBA and NSL-UEBA datasets, respectively, outperforming existing methods in terms of precision and resilience.


{\renewcommand{\baselinestretch}{1}\normalsize
\printbibliography
}

\begin{IEEEbiographynophoto}{Zilin Huang}
\renewcommand{\baselinestretch}{1}\selectfont
is currently an undergraduate student at the School of Cyberspace Security(School of Cryptology), Hainan University. Her research interests include AI security and big data.
\end{IEEEbiographynophoto}

\vspace{-2cm}

\begin{IEEEbiographynophoto}{Xiulai Li}
\renewcommand{\baselinestretch}{1}\selectfont
(Member, IEEE) received his B.S. degree in Electronic Information Science and Technology from Beijing Forestry University, Beijing, China, in 2015. He received his M.E. degree in Software Engineering from Hainan University, Haikou, Hainan, China, in 2019. He was the CEO of Hainan Hairui ZhongChuang Technology Co. Ltd. Now, He is studying for a Ph.D in Cyberspace Security at Hainan University, Haikou, Hainan, China. His current research interest includes artificial intelligence and security.
\end{IEEEbiographynophoto}

\vspace{-2cm}

\begin{IEEEbiographynophoto}{Xinyi Cao}
\renewcommand{\baselinestretch}{1}\selectfont
is currently an undergraduate student at the School of Computer Science and Technology, Hainan University. Her research interests include computer vision and big data.
\end{IEEEbiographynophoto}

\vspace{-2cm}

\begin{IEEEbiographynophoto}{Ke Chen}
\renewcommand{\baselinestretch}{1}\selectfont
is currently an undergraduate student at the School of Cyberspace Security, Hainan University. Her research interests include blockchain and cyber security.
\end{IEEEbiographynophoto}

\vspace{-2cm}

\begin{IEEEbiographynophoto}{Longjuan Wang}
\renewcommand{\baselinestretch}{1}\selectfont
received her B.S. degree in Beihang University, Beijing, China, in 1999. She received her M.E. degree in Huazhong University of Science and Technology, Wuhan, Hubei, China, in 2005. She is studying for a Ph.D in Cyberspace Security at Hainan University, Haikou, Hainan, China. She is the lecturer of Hainan University. Her current research interest includes artificial intelligence and security.
\end{IEEEbiographynophoto}

\vspace{-2cm}

\begin{IEEEbiographynophoto}{Logan Bo-Yee Liu}
\renewcommand{\baselinestretch}{1}\selectfont
(Member, IEEE) received his B.S. degree in Network engineering from Hainan University, Haikou, China. He received his Mphil degree in PAMI from The University of Chinese Academy of Science, Beijing, China. His research interests include Edge AI and Robotics.
\end{IEEEbiographynophoto}

\end{document}